\documentclass{aa}  

\usepackage{graphicx}
\usepackage{txfonts}
\usepackage{caption}
\titlerunning{The impact of rotation on the pulsation modes of V1790 Ori}
\authorrunning{X.Y Sun \& A.G Hernández \& and Z.Y Zuo et al}

\begin{document} 

   \title{Forward Modeling of the $\delta$ Sct Star V1790 Ori: $\Delta \nu$, $\Omega$, Resolution and Non-adiabatic Effects}

   \author{Xiaoya Sun\inst{1,2},
          Antonio García Hernández\inst{2},
          Zhaoyu Zuo\inst{1,3},
          Juan Carlos Suárez\inst{2},
          Yifan Wang\inst{4},
          Ruixuan Tang\inst{5},
          Giovanni M. Mirouh\inst{2},
          Taozhi Yang\inst{1}
          }

   \institute{Ministry of Education Key Laboratory for Nonequilibrium Synthesis and Modulation of Condensed Matter, School of Physics, Xi'an Jiaotong University, Xi'an 710049, People's Republic of China\\
              \email{zuozyu@xjtu.edu.cn}
         \and
             Departamento de Física Teórica y del Cosmos, Universidad de Granada, Campus de Fuentenueva s/n, E-18071, Granada, Spain\\
             \email{agh@ugr.es}
             \and 
             Institut f\"{u}r Astronomie und Astrophysik, Kepler Center for Astro and Particle Physics, Eberhard Karls, Universit\"{a}t, Sand 1, D-72076
T\"{u}bingen, Germany
             \and
             Department of Mathematical Sciences, University of Liverpool, Liverpool L69 3BX, UK\
             \and
             School of Astronomy and Space Science, University of Science and Technology of China, 230026 Hefei, Anhui, PR China.\
             }

   \date{Received ; accepted }
 
  \abstract 
   {}
   {We investigate the role of the large separation, rotational correction order, structural resolution, and non-adiabatic effects in the forward modelling of the rotating $\delta$ Scuti star V1790 Ori.}
   {We extract 69 significant frequencies from TESS data and determine $\Delta\nu \simeq 82$ $\mu$Hz. Rotating MESA models are computed at low and high mesh resolution. Their pulsation frequencies are calculated with GYRE, in both adiabatic and non-adiabatic form with first-order rotational corrections, and with FILOU, including second-order rotational corrections.}
   {Using $\Delta \nu$ as a structural constraint is necessary to reduce model degeneracy in V1790 Ori and to obtain physically plausible mode identifications. For the selected minimum-misfit reference model, and considering only the 40 modes with consistent $(n,\ell,m)$ labels across all configurations, the RMS$_{40}$ theoretical frequency differences are 0.442 $\mu$Hz between low and high spatial resolution, 0.062 $\mu$Hz between adiabatic and non-adiabatic calculations, and 2.962 $\mu$Hz between GYRE and FILOU calculations. When all 48 fitted frequencies are included, the corresponding RMS$_{48}$ values increase to 1.033, 2.326, and 3.931 $\mu$Hz, respectively. Relative to the observed frequencies, increasing the spatial resolution slightly reduces the RMS residuals, from 4.457 to 4.387 $\mu$Hz for RMS$_{40}$ and from 4.715 to 4.682 $\mu$Hz for RMS$_{48}$. Including non-adiabatic effects changes the residuals only marginally, from 4.387 to 4.381 $\mu$Hz for RMS$_{40}$ and from 4.682 to 4.673 $\mu$Hz for RMS$_{48}$. The largest change is obtained when the FILOU calculation, which includes second-order rotational corrections, is compared with the GYRE calculation: the residuals increase from 4.387 to 5.331 $\mu$Hz for RMS$_{40}$ and from 4.682 to 5.270 $\mu$Hz for RMS$_{48}$.}   
   {The comparison between RMS$_{40}$ and RMS$_{48}$ shows that the 40 stable-label modes are useful for isolating direct frequency shifts caused by different modelling choices, whereas the full 48-frequency comparison additionally captures the effect of modes with uncertain identifications. For the stable-label subset, increasing spatial resolution has a small but measurable effect, and non-adiabatic effects have only a very limited direct impact on the theoretical frequencies. However, the larger RMS$_{48}$ value in the adiabatic -- nonadiabatic comparison shows that non-adiabatic effects can still affect the closest-mode matching or mode identification for a small subset of frequencies. Second-order rotational corrections produce the largest changes in the theoretical frequencies, but their ability to improve agreement with observations should be reassessed through a denser model grid and a self-consistent FILOU optimisation. Notably, the 260.672 $\mu$Hz peak—previously identified as the fundamental radial mode—shows uncertain mode identification and requires further systematic analysis. The present results should therefore be interpreted mainly as a diagnostic of modelling systematics and mode-identification robustness, rather than as a definitive selection of a unique global mode identification solution.}  


   \maketitle

\section{Introduction}

Stellar oscillations provide a powerful window into the internal structures of stars. Asteroseismology, the study of stellar pulsations, has emerged as a key technique for probing stellar interiors with unprecedented precision \citep{Aerts2021}. Among the various classes of pulsation stars, $\delta$ Scuti stars occupy a unique position on the Hertzsprung-Russell (HR) diagram, located in the classical Cepheid instability strip where it crosses the main sequence (MS) \citep{Breger2000,Uytterhoeven2011,Holdsworth2014,Guzik2021}. These intermediate-mass (1.5-2.5 $M_{\odot}$) variables show both radial and non-radial modes \citep{Breger2000,Qian2018,Murphy2020,Sun2021,Yang2025}, usually identified as low-radial-order low-angular-degree pressure ($p$) modes \citep{Uytterhoeven2011,Holdsworth2014,Guzik2018}. Moreover, \cite{Uytterhoeven2011} detected gravity ($g$) modes in many $\delta$ Scuti stars.

Thanks to the high precision and long-duration continuous light curves provided by space telescope missions such as CoRoT \citep{Baglin2006}, Kepler \citep{Borucki2010} and TESS \citep{Ricker2014}, we know that many $\delta$ Scuti stars exhibit regular frequency patterns in their rich pulsation spectra \citep{AGH2009,AGH2013,Paparo2016a,Paparo2016b,Bedding2020,Yang2021,Sun2023}. Recently, numerous studies have investigated these complex spectra from various perspectives, such as the low-order large frequency separation ($\Delta \nu$, the frequency spacing between consecutive overtones of the same angular degree, \citealt{{AGH2015,AGH2017,Suarez2014,Rodriguez-Martin2020,Bedding2020}}), rotational splitting \citep{Ram2021,Guo2024}, and (near) equidistant frequency patterns such as triplet or quintuplet structures \citep{Ram2021,Sun2023}. Moreover, the large frequency separation is closely related to the star’s mean density and provides crucial constraints on stellar models \citep{Reese2008,Suarez2014,AGH2015,AGH2017}.

However, stellar rotation is a ubiquitous and crucial physical process. $\delta$ Scuti stars can rotate as fast as $\sim$ 250 km/s, with a mean at $v$sin$i$ $\sim$ 150 km/s for A-type stars \citep{Royer2009}. Rotation affects both their structure and pulsations: the centrifugal force causes significant deformation of the star \citep{Monnier2010,Kippenhahn2013}, while the Coriolis force modifies pulsation frequencies and obfuscates regular patterns in the spectra \citep{Lignieres2006}. Several studies have shown that stellar rotation leads to significant modifications to pulsation frequencies, with the impact becoming increasingly pronounced as higher-order rotational effects, such as second-order terms and non-degeneracy effects, are incorporated \citep{Reese2006,Aerts2010,Ballot2013}. \cite{Guo2024} found that the theoretical splitting asymmetry predicted by the second-order effect induced by rotational aspherical distortion can quantitatively explain the observed asymmetry.

Moreover, \cite{Reese2006} showed that for a star with a mass of $M$ = 1.9 $M_{\odot}$ and a radius of $R$ = 2.3 $R_{\odot}$, perturbative methods become invalid at CoRoT-level precision when $v$sin$i$ exceeds 50 or 75 km/s. At higher rotation rates (e.g. 0.59 $\Omega_{K}$, $\Omega_{K}$ is the Keplerian break-up rotation rate), perturbative frequency spectra differ significantly from fully non-perturbative results, potentially leading to incorrect interpretations. Non-perturbative calculations showed a re-shaping of mode geometries, which differ significantly from spherical harmonics. Fortunately, the large separation is still present and its relation with the stellar mean density is the same as for the moderate-and low-rotation cases \citep{AGH2015,Mirouh2019}.

However, even with fully non-perturbative calculations, the reliability of predicted frequencies depends not only on the physical assumptions but also on the numerical resolution of stellar models. As observational precision improves, modelling uncertainties become increasingly important, particularly those arising from insufficient structural resolution in models \citep{Moya2008,Li2025}, which can introduce systematic errors in the equilibrium structure and propagate directly into the oscillation frequencies \citep{Li2025}.

Furthermore, oscillation calculations often assume adiabatic pulsations. In reality, energy exchanges such as radiative damping and convective interactions occur. Theoretical photometric amplitude ratios and phase differences are highly sensitive to non-adiabatic effects \citep{Moya2003,Montalban2007}, which makes it essential to include non-adiabatic effects when interpreting high precision asteroseismic data.

In this work, we select V1790 Ori from \cite{Bedding2020} as our target star. Its frequencies exhibit some regularity, likely related to the large separation and therefore to the mean stellar density, and it has not yet been modelled in detail. With a moderate rotation velocity of 70 km/s, V1790 Ori provides an ideal case for studying the impact of rotation on pulsation modes. We compute rotational MESA models at both low and high resolution and compare observed frequencies with theoretical ones from various pulsation codes, including adiabatic, non-adiabatic and second-order rotational corrections. 

This paper is organized as follows. In Sect. 2, we introduce physical parameters of V1790 Ori, observational data and frequency analysis. In Sect. 3, we determine the large separation parameter. Then, we describe in detail the setup of the model grid and the pulsation codes used in our analysis (Sect. 4 and Sect. 5). Moreover, we explore the role of large separation, $\Delta \nu$, in mode identification and stellar model constraints. We investigate the impact of first vs second order rotational corrections, low vs high spatial resolutions, and adiabatic vs non-adiabatic effects on the pulsation modes (Sect. 6 and Sect. 7). In Sect. 8, we summarize the overall results of this work.

\section{V1790 Ori}

   V1790 Ori was identified as a young $\lambda$ Boo star by \cite{Gray1993}. Later, \cite{Andrievsky2002} confirmed the membership of this star in the $\lambda$ Boo
   class, with normal C and O, and [\text{Fe/H}]= -1.0 dex. They estimated the surface rotational velocity ($v$sin$i$) to be 70 km/s. \cite{Murphy2015} obtained a new spectrum of this star and classified it as A7 V kA2mA2 $\lambda$ Boo. Recently, 
   \citet{Bedding2020} discovered V1790 Ori is a $\delta$ Scuti star with regular frequency patterns. \citet{Bedding2020} and \citet{Murphy2023} identified the peak at 260.764 $\mu$Hz as the fundamental radial mode. The parameters of V1790 Ori are listed in Table~\ref{parameter}.

\begin{table}
\caption{\label{parameter}Stellar Parameters of V1790 Ori.}
\centering
\begin{tabular}{ccccc}
\hline\hline
$T_{\mathrm{eff}}$ & $L$ & $v$sin$i$ & $\Delta$$\nu$ & Reference\\ 
(K) & ($L_{\mathrm{\odot}}$) & (km/s) & ($\mu$Hz) &  \\
\hline
   8000 $\pm$ 250 & - & 70 $\pm$ 10 & - & a \\  
   7745 $\pm$ 90 & 7.94 $\pm$ 2.21 & 70 & - &  b \\
   8780 $\pm$ 176 & 13.21 $\pm$ 0.98 & - & 87 & c \\
   - & - & - & 89 & d \\
\hline
\end{tabular}
\tablefoot{$\Delta \nu$: the large separation; a: \cite{Andrievsky2002}; b: \cite{Paunzen2002}; c: \cite{Bedding2020}; d: \cite{Murphy2023}.}
\end{table}

   In this work, light curves from TESS are downloaded from the Mikulski Archive for Space Telescopes (MAST) \footnote{https://archive.st sci.edu/}. We use Pre-Search Data Conditioning Simple Aperture Photometry (PDCSAP) to calculate the Fourier amplitude spectra. V1790 Ori was observed in Sector 6 in 2-min cadence mode for a total of 21.7 day, so the Nyquist frequency is 4167 $\mu$Hz.
   
   We extract the frequency content of V1790 Ori using MultiModes \citep{Pamos2022}, which is a Python routine designed to calculate the Fast Lomb-Scargle periodogram \citep{Press1989} of light curves. It extracts significant signals one by one and performs non-linear optimization to fit their corresponding frequencies, amplitudes, and phases, modelling the entire signal as a multi-sine function.
   
   The fitted components are then subtracted from the original signal, and the algorithm restarts the loop with the residuals as the new input. This process is repeated until the stop criterion (signal to noise ratio > 4.0, \citealt{Breger1993}) is reached. The code is also capable of filtering out suspicious spurious frequencies, low-amplitude signals above the Rayleigh resolution, and possible combination modes. A detailed description of the code is provided in \cite{Pamos2022} and in its public GitHub repository \footnote{https://github.com/davidpamos/MultiModes}.
   
\begin{figure}
   \centering
   \includegraphics[width=\hsize]{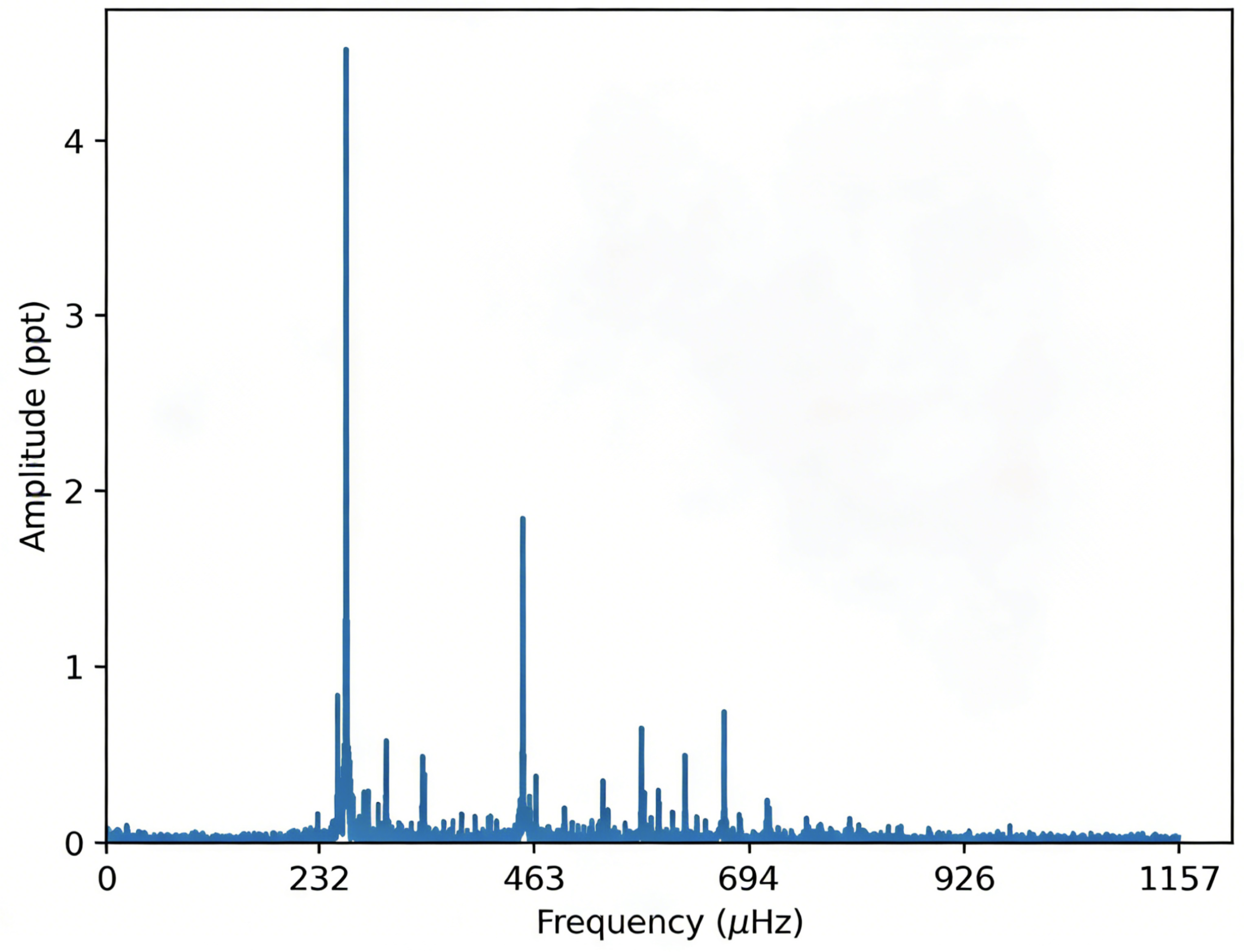}
      \caption{Frequency spectrum of V1790 Ori.}
         \label{LS}
   \end{figure}

   In total, 69 frequencies are extracted for V1790 Ori. The amplitude spectrum of the single sector (S6) of TESS data is shown in Fig.~\ref{LS}, and the frequencies along with their amplitudes and S/N (signal to noise ratio) are listed in Table~\ref{appendix}.

\section{Seismic index: large separation}

In this work, we determine the large separation $\Delta\nu$ of V1790 Ori, using the methodology provided by \cite{AGH2009,AGH2013} and \cite{Ram2021}. We compute the Fourier transform (FT), the autocorrelation function (AC), and the histogram of frequency differences (HFD) to the 30 highest-amplitude frequencies above 57.87 $\mu$Hz, avoiding $g$ modes. The bin size of HFD is 0.5 $\mu$Hz, the Rayleigh resolution of AC is 0.5 $\mu$Hz. To avoid the blurring of the échelle diagram caused by the large dynamic range of the amplitudes and to facilitate a clearer visualization of the frequency distribution, the amplitudes of frequencies have been normalized to unity when computing the transformations. If the signal contains a periodicity resembling a Dirac comb, the AC will show a peak at $\Delta\nu$ and additional peaks at its multiples (2$\Delta\nu$, 3$\Delta\nu$, etc.). In contrast, the FT produces a peak at $\Delta\nu$ and its submultiples ($\Delta\nu$/2, $\Delta\nu$/3, etc.) due to the nature of the transformation. 

\begin{figure*}
   \includegraphics[width=\textwidth]{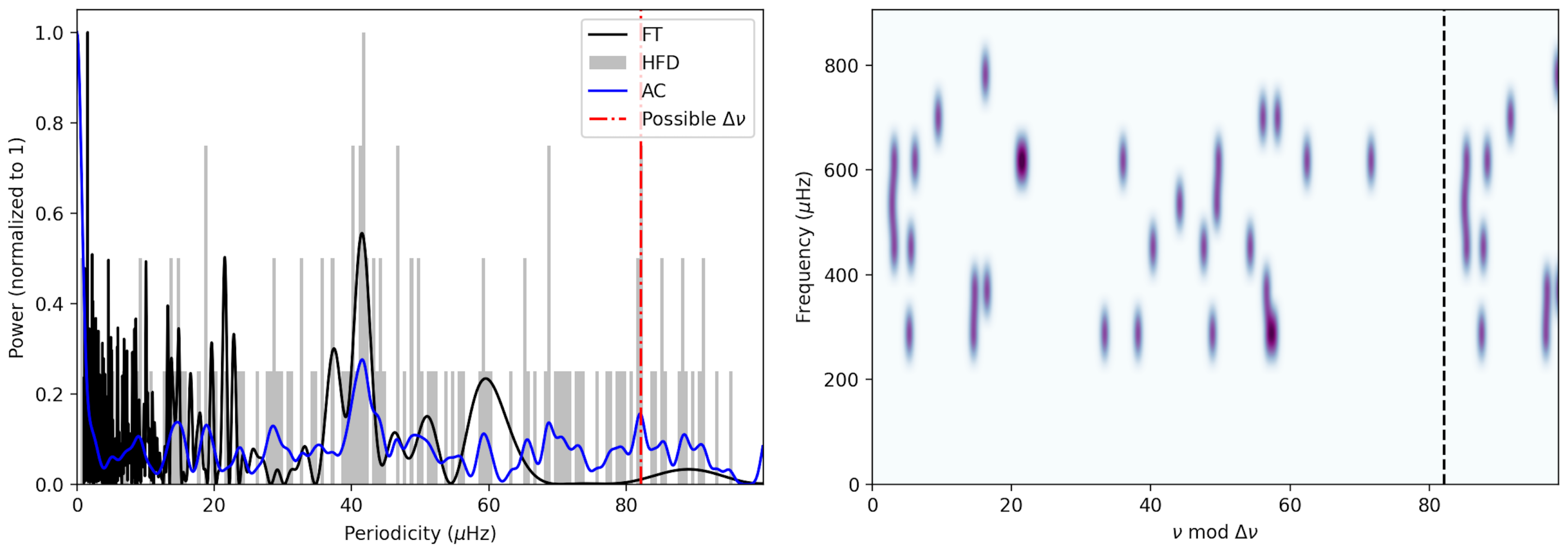}
      \caption{In the left panel, the black, gray and blue lines represent the Fourier transform (FT), the histogram of frequency differences (HFD), and the autocorrelation function (AC), respectively, using the methodology of \cite{AGH2013} and \cite{Ram2021}. The red dotted-dashed line indicates the possible value of large separation $\Delta\nu$ ($\sim$ 82 $\mu$Hz). The right panel shows the échelle diagram of the frequencies. The value of $\Delta\nu$ = 82 $\mu$Hz is used for the plot.}
         \label{V1790ED}
   \end{figure*}

The result is shown in Fig.~\ref{V1790ED}. In the left panel, the AC and HFD show a peak around 82 $\mu$Hz. However, in the FT, no peak is observed around $\Delta\nu$, so its identification relies on the presence of at least one of its submultiples. The FT, AC and HFD show a peak around 41 $\mu$Hz, which may be half the value of the large separation. In addition, we test the detection of $\Delta\nu$ using different subsets containing 20, 40, 50, 60, and 69 frequencies. We find that the 1/2 $\Delta\nu$ ($\sim$ 41 $\mu$Hz) signal persists in all subsets and remains highly significant. Moreover, in the right panel, the échelle diagram shows the alignment of several frequencies when 82 $\mu$Hz is chosen as the large separation. Thus, we consider 82 $\mu$Hz as $\Delta\nu$ of V1790 Ori, being the most common value obtained by different diagnostic techniques.

\section{Stellar evolutionary models}
\subsection{Grid input parameters}

In Sects. 2 and 3, we use high precision observational data to extract the oscillation frequencies of V1790 Ori and it's characteristic patterns, the large separation. These quantities not only trace the mean density and internal stratification of the star, but also provide direct observational constraints for stellar modelling. To further interpret these seismic features in a physical context, we construct in this section a grid of stellar evolutionary models spanning a range of initial parameters.

Stellar evolutionary models are calculated using Modules for Experiments in Stellar Astrophysics (MESA r24.08.1; \citealt{Paxton2011,Paxton2013,Paxton2015,Paxton2018,Paxton2019,Jermyn2023}). In the grid, we select the mass ($M$) range of 1.50 - 2.00 $M_{\mathrm{\odot}}$ with a step size of 0.01 $M_{\mathrm{\odot}}$. This range not only covers the mass of V1790 Ori (1.55 $\pm$ 0.01 $M_{\mathrm{\odot}}$) reported by \cite{Panda2024} but also falls within the typical mass range of $\delta$ Scuti stars. The metallicity $Z$ ranges from 0.005 to 0.020 with a step of 0.005.

In our models, convection is treated using the mixing-length theory (MLT), which simplifies the complexity of turbulent convection into a parametrized framework \citep{Joyce2023}. Several studies have investigated the value of the $\alpha_{\mathrm{MLT}}$ parameter for $\delta$ Scuti stars and have provided corresponding constraints: \cite{JDD2023a} found $\alpha_{\mathrm{MLT}}$ $\approx$ 1.0 for SX Phoenicis (a population II star). \cite{Miszuda2022} studied an eclipsing binary system with a $\delta$ Scuti component, AB Cas, and constrained $\alpha_{\mathrm{MLT}}$ to be within [1.2, 1.5]. In addition, some studies have reported that $\alpha_{\mathrm{MLT}}$ of $\delta$ Scuti stars is typically less than 1.0 \citep{JDD2021,JDD2023b,Sun2024}. Finally, we set $\alpha_{\mathrm{MLT}}$ from 1.0 to 1.8 in increments of 0.2.

In MESA, stellar rotation is treated using the one-dimensional shellular approximation, assuming constant rotation on isobaric surfaces (\citealt{Zahn1992}; \citealt{Meynet1997}). Rotational effects are included in 1D models via centrifugal acceleration and structural deformation (\citealt{Kippenhahn1970}; \citealt{Paxton2013,Paxton2019}). During pre-main sequence (pre-MS) contraction, the surface rotational velocity increases; for instance, a 1.7 $M_{\odot}$, solar metallicity model with an initial velocity of 10 km/s reaches 142.6 km/s at the zero-age main sequence \citep{Gautam2025}. In our grid, rotation is applied from the pre-MS stage with solid-body profiles and initial surface rotational velocities of 8.6 – 9.6 km/s (0.5 km/s steps) to achieve a target surface velocity of 70 km/s (\citealt{Andrievsky2002}; \citealt{Paunzen2002}). All models are evolved until the central H1 fraction drops below $10^{-10}$.

\subsection{Structural resolution}

In MESA, structural resolution refers to the number of radial (Eulerian) or mass (Lagrangian) grid points used in the calculation of stellar profiles, which is also called the mesh resolution. By default, MESA employs an adaptive mesh refinement (AMR), meaning that the model is “remeshed” or adaptively adjusted between time steps (\citealt{Paxton2011}). 

The structural resolution of stellar models is controlled by the "mesh$\_$delta$\_$coeff" parameter in MESA. This parameter does not directly set the mesh resolution itself but acts as a multiplier on the default mesh resolution. By adjusting this parameter, users can specify how much more (or less) the model resolution should be relative to MESA’s default AMR resolution. For example, when "mesh$\_$delta$\_$coeff = 1.0", the model uses MESA’s default structural resolution; when "mesh$\_$delta$\_$coeff = 0.1", the resolution is around ten times finer than the default \citep{Li2025}.

In this work, we compute stellar models at two different structural resolutions: one using MESA’s default structural resolution ("mesh$\_$delta$\_$coeff = 1.0", low resolution), and the other using mesh points $\sim$ 4000 as recommended by \cite{Moya2008} (i.e., "mesh$\_$delta$\_$coeff = 0.2", high resolution).

\section{Stellar pulsation calculations}

Based on the model grid established in Sect. 4, we then compute the theoretical pulsation frequencies for all models. We focus on $p$ modes and neglect $g$ modes. 

We use GYRE v8.0 \citep{Townsend2013} to calculate adiabatic pulsation modes and use the second-order Gauss-Legendre Magnus difference scheme (MAGNUS$\_$GL2) for models. We perform frequency scans from 200 to 1000 $\mu$Hz using a regular frequency grid of 500 points for $p$ modes \citep{Gautam2025}.

GYRE calculates the adiabatic frequencies with lowest-order rotational correction: a kinematic Doppler shift arising from the transformation between the inertial and co-rotating frames, and first-order dynamical perturbations caused by the Coriolis force \citep{Townsend2018,Goldstein2020,Gautam2025}. Together, these constitute the first-order rotational correction. However, \cite{Ballot2013} found that the validity range of the first-order correction is approximately limited by $\Omega_{\mathrm{V}} = \Omega_{\mathrm{K}} \sqrt{2\,\delta\omega / \omega}$ ($\Omega_{\mathrm{V}}$ is the upper rotation rate up to which the approximation remains valid, $\Omega_{\mathrm{K}}$ is the Keplerian break-up rotation rate, $\delta \omega$ is the tolerated frequency error scale and $\omega$ is the pulsation frequency). Therefore, we use FILOU to compute adiabatic oscillation frequencies including second-order rotational corrections.

FILOU \citep{Suarez2008} is a pulsation code designed to compute both radial and non-radial stellar oscillation frequencies, using a perturbative theory. The code applies rotational approximation to the oscillation frequencies up to second order, including near-degeneracy effects and stellar structure deformation caused by the centrifugal force. Additionally, FILOU can handle either uniform or shellular rotation profiles \citep{Suarez2006b,Suarez2008}. Moreover, \cite{Ballot2013} found that, when second-order near-degeneracy effects are taken into account, the high-frequency modes of solar-type stars can be successfully reproduced up to $\Omega \approx 0.09 \Omega_{\mathrm{K}}$. 

In addition, we also consider that non-adiabatic effects (accounting for energy exchanges such as radiative losses and convective transport) can provide a better description of the oscillations. And non-adiabatic effect has a direct impact on the value of the mode frequencies \citep{Grigahcene2005}. Therefore, we use GYRE v8.0 to perform non-adiabatic calculations with the adiabatic method. 

In Table \ref{grid_fre}, we summarize the different combinations of model parameters described in Section 4.1, along with the pulsation properties presented in Section 5, in order to clearly show the relationship between each model parameter combination and the corresponding pulsation characteristics.

\begin{table*}
\caption{\label{grid_fre}Summary of stellar model parameter combinations and their corresponding pulsation properties.}
\centering
\begin{tabular}{c c c c c c c c c c c c c c c}
\hline\hline

Parameter & Range/Method & Step/Rotational Correction & Note \\ 
\hline
$M$ & 1.5 -- 2.0 $M_{\odot}$ & 0.01 $M_{\odot}$ &  Mass  \\
$Z$ & 0.005 -- 0.020 & 0.005 &  Metallicity  \\
$\alpha_{\rm MLT}$ & 1.0 -- 1.8 & 0.2 &  Mixing-length Parameter \\
$v_{\rm rot, initial}$ & 8.6 -- 9.6 km/s & 0.5 km/s & Initial Surface Rotational Velocity   \\
GYRE & Adiabatic / Non-adiabatic Frequencies & 1st-order Rotational Correction &  Applied to Full Model Grid \\
FILOU & Adiabatic Frequencies & 2nd-order Rotational Correction &  Applied to Full Model Grid \\
\hline
\end{tabular}
\tablefoot{The full model grid is constructed from different combinations of $M$, $Z$, $\alpha_{\rm MLT}$, and $v_{\rm rot, initial}$.}
\end{table*}

With the observational constraints and theoretical tools established, we now proceed to select the minimum-misfit stellar model and quantify the impact of different physical and numerical treatments.

\section{Results}
\subsection{The critical role of $\Delta \nu$ in model selection}

To select candidate models for further analysis, we firstly use the following observational constraints: effective temperature, luminosity, stellar surface rotational velocity. Then we compare the observed and theoretical frequencies following Equation (3) of \cite{Chen2019}, i.e., $\chi^2 = \frac{1}{k} \sum (|\nu_{\rm obs,i} - \nu_{\rm mod,i}|^2$), and identify the minimum-misfit model by minimizing the corresponding $\chi^2$ value. In this work, we adopt the $\chi^2$ as a relative misfit metric to rank our models in a homogeneous way to study trends as a function of the number of fitted frequencies.

We select models with effective temperature in the range of 7600 K < $T_{\mathrm{eff}}$ < 9000 K to cover the values of V1790 Ori given by \cite{Andrievsky2002}, \cite{Paunzen2002} and \cite{Bedding2020}. We derive stellar luminosity from $Gaia$ parallax by first computing the absolute G-band magnitude $M_{\mathrm{G}}$, $-2.5 \log_{10} L = M_{\mathrm{G}} + BC_{\mathrm{G}}(T_{\mathrm{eff}}) - M_{\mathrm{bol\odot}}$ \citep{Andrae2018}, and then applying the bolometric correction $BC_{\mathrm{G}}(T_{\mathrm{eff}}$) through $M_{\mathrm{G}} = G - 5 \log_{10} r + 5 - A_{\mathrm{G}}$ \citep{Andrae2018}. Thus, the luminosity range is 0.93 < $\log_{10} L$ < 1.13. According to \cite{Andrievsky2002} and \cite{Paunzen2002}, and to save computational resources, we adopt a surface rotation velocity range of 69 -- 71 km/s. It should be noted that the adopted surface rotation velocity range is derived from the observed $v$sin$i$ values and therefore assumes an inclination angle of $i$ = 90$^\circ$, i.e. $v_{\rm eq}$ $\approx$ $v$sin$i$. Since the inclination angle is not independently constrained, the true equatorial velocity may be higher than the adopted value. This uncertainty may affect the subsequent predictions of rotational splittings, the mode identification, and the comparison between the GYRE and FILOU results.

Because $\delta$ Scuti stars exhibit closeness of mode frequencies between different models, accurately identifying their observed modes is challenging. Therefore, in the frequency $\chi^2$ formula, we don't assign any possible mode identifications to the peaks (i.e., we don't consider their radial order $n$ and spherical degree $\ell$), but instead compare the observed frequencies to their nearest model frequencies. A one-to-one correspondence between observed and theoretical frequencies is enforced, ensuring that each theoretical mode is assigned to at most one observed frequency. We progressively increase the number of fitted modes (from 1 to 69, ordered by decreasing amplitude) to calculate $\chi^2$ for all models and search for the one with the minimum value. Then we repeat the entire fitting process for different configuration combinations (i.e., different combinations of resolutions, rotational corrections, and non-adiabatic effects). In order to investigate frequency differences among various configuration combinations, we need to select a single minimum-misfit model across all combinations, thereby reducing the influence of other effects.

We then examine how the normalized merit function, $\chi^2$/N, varies with the number of fitted frequencies, N, for the different combinations of pulsation code, physical assumption, and resolution. For each given value of N, the minimum-misfit model is identified within each configuration as the model that minimizes $\chi^2$; this minimum value is then normalized by N and plotted as $\chi^2$/N. The resulting curves therefore trace how the best achievable fit quality changes as an increasing number of observed frequencies is used to constrain the models.

It should be emphasized that, because the matching between observed and theoretical frequencies is model dependent and the mode identification is not fixed a priori, the $\chi^2$ values are not interpreted in a probabilistic sense in this work. Consequently, the absolute value of $\chi^2$ is not used to assign statistical confidence levels. Instead, the physically relevant information is taken to be the stability of the low-misfit regions and the robustness of the corresponding parameter ranges as N increases. This approach allows us to assess which configuration provides a stable and reliable fit when progressively more frequencies are included.

   \begin{figure}
   \centering
   \includegraphics[width=\hsize]{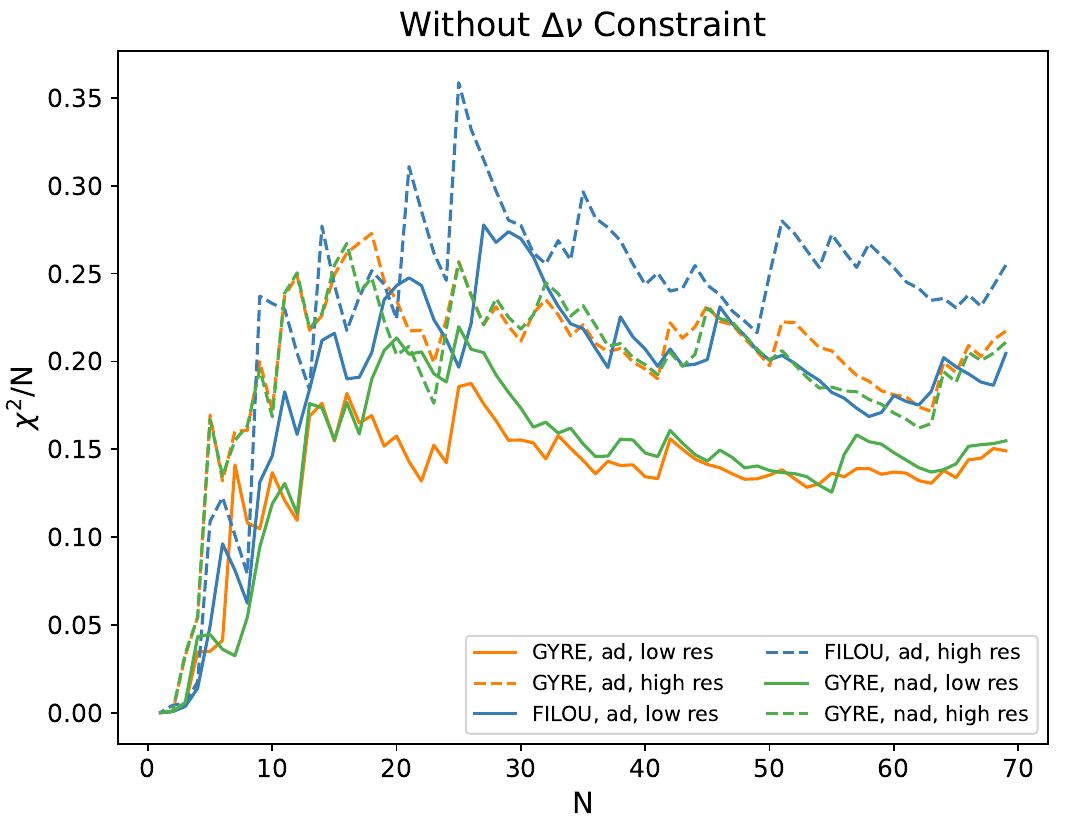}
      \caption{Variation of the normalized merit function, $\chi^2$/N, with the number of fitted frequencies, N, for different pulsation code, non-adiabatic effect, and spatial resolution configurations. The frequencies are added sequentially from 1 to 69 after being sorted in descending order of amplitude; their detailed values are listed in Table~\ref{appendix}. For each given N, the minimum-misfit model is identified within each configuration, namely the model with the minimum $\chi^2$, and this minimum value is then normalized by N. Thus, the curves show how the best achievable fit quality and its stability vary as more frequencies are included in the fitting. Orange, blue, and green curves correspond to GYRE adiabatic (ad), FILOU adiabatic, and GYRE non-adiabatic (nad) pulsation calculations, respectively. Solid lines denote low spatial resolution, whereas dashed lines denote high spatial resolution.}
         \label{chi2_n_without_dnu}
   \end{figure}

\subsubsection{Model selection without $\Delta \nu$ constraint}

In the first step, we don't use the large separation $\Delta \nu$ to constrain models. Therefore, when selecting candidate models, we use parameters $T_{\mathrm{eff}}$, $L$, $v_{\mathrm{rot}}$ as described above. Fig.~\ref{chi2_n_without_dnu} shows the variation of $\chi^2$/N with the number of fitted frequencies in different configuration combinations. 

A comparison of the six curves reveals clear differences among the various configuration combinations as the number of fitted frequencies increases. At small N, all curves show pronounced variations in $\chi^2$/N, and the relative performance of the different configurations is not stable. This indicates that, when only a limited number of frequencies is included in the fit, the resulting best-fitting models are strongly affected by the particular subset of frequency constraints. Consequently, the corresponding $\chi^2$/N values are highly sensitive to N, and this regime is not suitable for determining the final number of fitted frequencies.

The effect of spatial resolution can also be assessed from the comparison between solid and dotted lines. The high-resolution results, shown by dotted lines, generally display larger fluctuations and, in most cases, higher values of $\chi^2$/N than their low-resolution counterparts. In particular, the FILOU-adiabatic-high resolution configuration remains at relatively high $\chi^2$/N over a broad range of N, with noticeable fluctuations, suggesting that its best-fitting models are more sensitive to the progressive inclusion of additional frequencies. The GYRE adiabatic and GYRE non-adiabatic high resolution configurations show comparable behaviour in some intervals, but they also exhibit substantial variations, especially in the range N = 15 – 35. Therefore, the high-resolution configurations do not necessarily yield the most stable solution for the subsequent analysis. A possible explanation is that increasing the mesh points scans better the stellar interior so small changes in the model configuration will translate in noticeable differences in the frequency modes. Thus, to properly account for these changes requires a very dense grid parameter space, maybe even modifying helium or particular metal abundances. Of course, this is not the objective of this work but it points to a way to go in the future.

From the point of view of the pulsation code and the non-adiabatic effect, the blue solid curve, corresponding to the FILOU-adiabatic-low resolution configuration, reaches relatively high values of $\chi^2/N$ in the intermediate N regime. This is particularly evident around N = 25 -- 35, where it lies noticeably above the other low resolution curves. The same number of modes is considered in the FILOU and GYRE comparisons, since both calculations include rotational splittings. Nevertheless, the FILOU frequencies are more affected by mode coupling and near-degeneracy. When coupled modes are involved, a change in the behaviour of one mode can also affect its coupled counterpart, and even models that are close in the grid or along the evolutionary sequence may yield noticeably different mode matches. This may explain why the FILOU configuration shows a less favourable and less stable normalized fit quality in this range as additional frequencies are included. The green solid curve, corresponding to the GYRE-nonadiabatic-low resolution configuration, performs relatively well overall and approaches the orange solid curve at large N. Nevertheless, it still shows appreciable fluctuations in some intervals and does not provide a systematically lower or more stable $\chi^2$/N than the orange curve.

By contrast, the orange solid curve, corresponding to the GYRE-adiabatic-low resolution configuration, shows lower and more stable values of $\chi^2$/N over the large N range. In particular, for N > 40, this curve remains among the lowest of all six configurations, indicating that the corresponding best-fitting models maintain a small normalized misfit as the number of fitted frequencies increases. Moreover, compared with the other curves, the orange solid curve does not show pronounced spike-like variations, suggesting a weaker sensitivity to the progressive increase in N and hence a more stable modelling behaviour.

All this reinforces the idea that we need high resolution and second-order rotational correction. However, it would require a much more detailed study. Nonetheless, to study the impact of without the large separation in the fit, it is better to use the most stable configuration, i.e., the GYRE-adiabatic-low resolution configuration.

After selecting this configuration, we further examined the behaviour of the orange solid curve as a function of N. In the interval N = 55 – 62, the variation of $\chi^2$/N is small and can be approximated by a slow, nearly linear trend. This suggests that adding further frequencies within this range does not substantially improve the normalized misfit. At the same time, no clear additional improvement is seen beyond this interval. We therefore identified N = 55 – 62 as a suitable range for the number of fitted frequencies.

To compare the fitting discrepancies consistently across different values of N, we take the number of degrees of freedom to be N and consider the normalised statistic $\chi^2$/N. Since $\chi^2$ follows a $\chi^2$ distribution with N degrees of freedom, $\chi^2$/N follows a Gamma(N/2,2/N)) distribution under the shape--scale parameterisation. Accordingly, the corresponding normalised critical value is obtained by dividing the original $\chi^2$ critical value by N.

According to \cite{Thompson1941}, for degrees of freedom greater than 30, the critical values of the $\chi^2$ distribution are tabulated at intervals of 10. At $p$ = 0.005, the tabulated values for N = 50, 60, and 70 are $\chi_{50,\,p=0.005}^2$ = 28.0, $\chi_{60,\,p=0.005}^2$ = 35.5, and $\chi_{70,\,p=0.005}^2$ = 43.3, corresponding to normalised values of 0.56, 0.59, and 0.62, respectively. Because the fits considered here involve intermediate values of N, the corresponding continuous critical boundary is calculated from the Gamma(N/2,2/N) distribution rather than read directly from the tabulated values.

For fits using N = 55 -- 62 frequencies, the resulting values of $\chi^2$/N range from 0.13 to 0.14, whereas the corresponding $p$ = 0.005 critical values range from 0.58 to 0.59. All fits lie well below their corresponding critical boundaries, indicating that the discrepancies between the theoretical and observed frequencies remain consistently low throughout the investigated range of N. Nevertheless, the final model is not selected solely on the basis of the minimum $\chi^2$/N. To retain a sufficiently large number of observational constraints while maintaining stable fitting quality, we adopt the fit using N = 60 frequencies. Thus, the model with parameters $M$ = 1.71 $M_{\mathrm{\odot}}$, $v_{\rm rot, initial}$ = 9.6 km/s, $Z$ = 0.01 and $\alpha_{\mathrm{MLT}}$ = 1.0 is identified as the minimum-misfit candidate.

\begin{figure}
   \centering
   \includegraphics[width=\hsize]{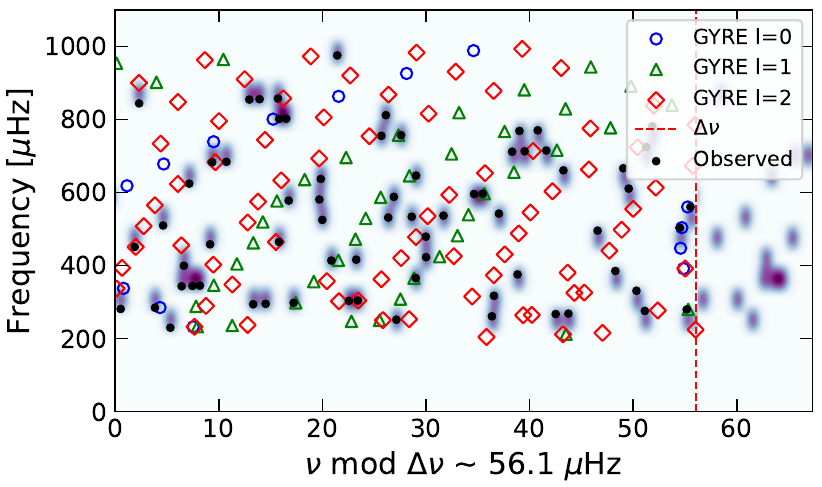}
      \caption{Échelle diagram of the minimum-misfit model obtained by fitting 60 observed frequencies (black dots) without large separation constraint. Model parameters: $M$ = 1.71 $M_{\mathrm{\odot}}$, $v_{\rm rot, initial}$ = 9.6 km/s, $Z$ = 0.01 and $\alpha_{\mathrm{MLT}}$ = 1.0. The blue, green, and red symbols denote theoretical modes with $\ell$ = 0, 1, and 2, respectively, computed using GYRE under the adiabatic approximation at low resolution. The purple symbols indicate the observed frequencies (normalized amplitudes). Red dashed line denotes the large separation $\sim$ 56.1 $\mu$Hz.}
         \label{echelle_without_dnu}
   \end{figure}

Fig.~\ref{echelle_without_dnu} shows the échelle diagram (as a check of structural consistency rather than as a fitting tool) of this candidate model, which gives a theoretical large separation of 56.1 $\mu$Hz. 

Following \cite{Suarez2014}, we compute the theoretical $\Delta \nu$ for consecutive oscillation modes of each model, defined as:

\begin{equation}
\Delta \nu_{\ell} = \nu_{n+1,\ell} - \nu_{n,\ell}  \label{equation3}
\end{equation}

where $\nu_{n,\ell}$ is the frequency of the mode with radial order $n$ (2 $\leq$ $n$ $\leq$ 8) and spherical degree $\ell$ = 0, 1, 2, assuming $m$ = 0. For each $\ell$, the actual $\Delta \nu_{\ell}$ is taken as the median of the individual values rather than the mean, because the regular pattern of the large separation can be blurred in the presence of avoided crossings, whereas the median is more robust against such outliers \citep{Suarez2014,Rodriguez-Martin2020}. The final theoretical large separation $\Delta \nu$ is the mean value of $\Delta \nu_{0}$, $\Delta \nu_{1}$ and $\Delta \nu_{2}$.

Fig.~\ref{echelle_without_dnu} shows the theoretical modes are distributed in a scattered manner without a clear regular pattern. This suggests that the presence of numerous pulsation modes, and blind frequency fitting may result in unreliable or misleading results and mode identification.

\subsubsection{Model selection with $\Delta \nu$ constraint}

In Sec. 6.1.1, we determine the minimum-misfit model using only local oscillation frequencies. Since blind fitting of individual frequencies yield high uncertainties due to the lack of a proper mode identification strategy, we introduce a global seismic parameter, the large separation $\Delta \nu$, as an additional constraint. $\Delta \nu$ is closely related to the mean density of the star and provides a global constraint on the stellar structure \citep{Suarez2014,AGH2015}, thereby enabling a combined global and local characterization of the stellar interior.

\begin{figure}
   \centering
   \includegraphics[width=\hsize]{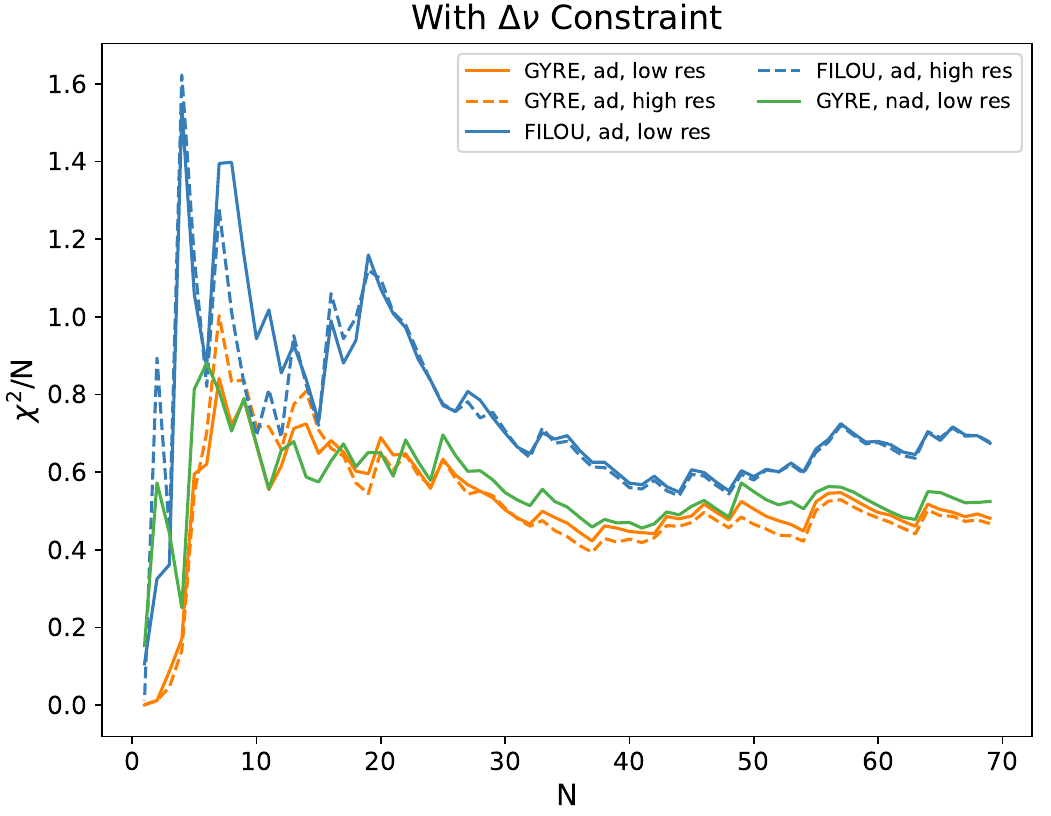}
      \caption{Same as in Fig.~\ref{chi2_n_without_dnu}, except that an additional constraint on the large separation $\Delta \nu$ is applied during the model selection process.}
         \label{chi2_n_dnu}
   \end{figure}

The results of the frequency fit, now including $\Delta \nu$ as a constraint on the best model, are presented in Fig.~\ref{chi2_n_dnu}. As shown in the figure, no candidate model is identified in the GYRE-nonadiabatic-high resolution configuration. This is because, for the adopted value of $\alpha_{\mathrm{MLT}}$, the large separation predicted by this particular model setup falls outside our predefined selection range.

It should be emphasized that the aim of introducing high spatial resolution model in the present work is to isolate the effect of increased spatial resolution on the pulsation frequencies, rather than to conduct a systematic exploration of the $\alpha_{\mathrm{MLT}}$ parameter space at high resolution. More generally, this work focuses on comparing different approaches to mode identification in a relatively simple $\delta$ Scuti star's amplitude spectrum, and on testing which strategy is required to obtain robust results (such as incorporating the large separation as a prior). Once we proved that, even this case need high resolution FILOU models (or StORM), then we can go for a correct mode identification doing a more dense grid but only using one code. For this reason, at high spatial resolution we considered only one value of $\alpha_{\mathrm{MLT}}$ in the MESA model calculations.

Moreover, the high resolution models are also computationally expensive in terms of both time and memory, which limited the generation of additional high resolution models with alternative $\alpha_{\mathrm{MLT}}$ values. A complete high resolution grid covering different $\alpha_{\mathrm{MLT}}$ values will be considered in future work to assess their effects on the fitting results more systematically.

Using the same analysis method as described in Sect. 6.1.1, we adopt adiabatic frequencies computed with GYRE at high resolution and determine N within the range 43 – 51. The resulting values of $\chi^2$/N range from 0.45 to 0.49, whereas the corresponding $p$ = 0.005 critical values range from 0.53 to 0.56. All fits lie below their respective critical boundaries, indicating that the discrepancies between the theoretical and observed frequencies remain low throughout the investigated range. Since the fits consistently remain within a stable low discrepancy regime, the final model is not selected solely on the basis of the minimum value of $\chi^2$/N, but also by considering the number of observational constraints included, numerical stability, and robustness with respect to variations in N. On this basis, we adopt the fit with N = 48 as the final model. The corresponding minimum-misfit model has parameters $M$ = 1.64 $M_{\mathrm{\odot}}$, $v_{\rm rot, initial}$ = 9.6 km/s, $Z$ = 0.01 and $\alpha_{\mathrm{MLT}}$ = 1.6.

\begin{figure}
   \centering
   \includegraphics[width=\hsize]{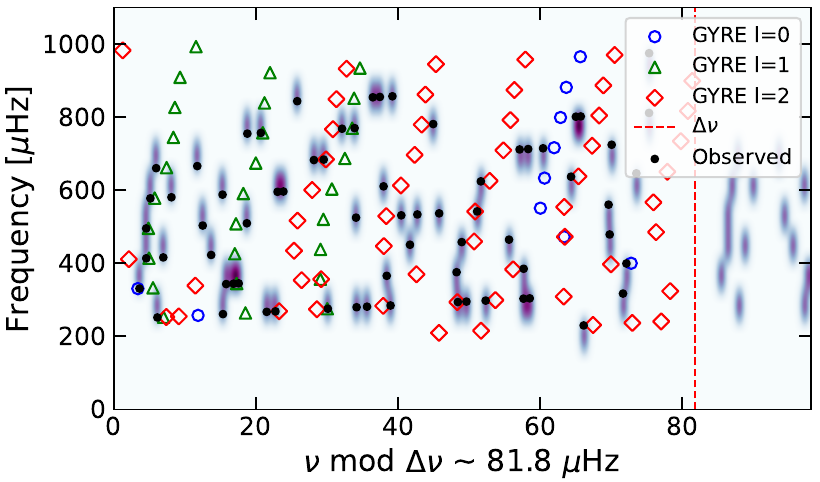}
      \caption{Échelle diagram of the minimum-misfit model obtained by fitting 48 observed frequencies (black dots) with large separation constraint. Model parameters: $M$ = 1.64 $M_{\mathrm{\odot}}$, $v_{\rm rot, initial}$ = 9.6 km/s, $Z$ = 0.01 and $\alpha_{\mathrm{MLT}}$ = 1.6. The symbols are same as in Fig.~\ref{echelle_without_dnu}, theoretical modes compute with GYRE in the adiabatic approximation at high resolution. Red dashed line denotes the large separation $\sim$ 81.8 $\mu$Hz.
}
         \label{echelle_dnu}
   \end{figure}

Fig.~\ref{echelle_dnu} shows the échelle diagram of the minimum-misfit model, which gives a theoretical large separation of $\sim$ 81.8 $\mu$Hz. This figure clearly shows the ridge structures for $\ell$ = 0, 1, 2. Compared with Fig.~\ref{echelle_without_dnu}, when numerous pulsation modes are present, introducing the large separation to guide frequency fitting is a more effective approach than blindly fitting frequencies, improving the accuracy of mode identification.  

Table~\ref{table:mmisfit-model} summarizes the parameters of the minimum-misfit models obtained for different configuration combinations using 48 observed frequencies and the large separation as constraints. A comparison of the results reveals that the inferred $M$, $\alpha_{\mathrm{MLT}}$, and $v_{\rm rot,initial}$ remain sensitive to the adopted pulsation treatment and numerical resolution, resulting in significant variations among the derived model parameters. These differences likely arise because the rotational correction, non-adiabatic effects, and spatial resolution introduce systematic shifts in the theoretical frequencies, which can move the minimum-misfit solution within the model parameter space. The derived parameters should therefore be interpreted as configuration dependent estimates. In future work, we will quantify these systematic effects using controlled numerical tests and construct a targeted high-resolution grid over the parameter ranges favoured by the present calculations.

\begin{table*}
\caption{Minimum-misfit model parameters for different configurations, obtained by fitting 48 frequencies with the $\Delta \nu$ constraint included.}             
\label{table:mmisfit-model}      
\centering          
\begin{tabular}{c c c c c}     
\hline\hline       
Configurations & $M$ ($M_{\mathrm{\odot}}$) & $Z$ & $\alpha_{\mathrm{MLT}}$ & $v_{\rm rot, initial}$ (km/s) \\ 
\hline                    
GYRE-adiabatic-low resolution & 1.62 & 0.01 & 1.4 & 9.1 \\  
GYRE-adiabatic-high resolution & 1.64 & 0.01 & 1.6 & 9.6 \\
FILOU-adiabatic-low resolution & 1.57 & 0.005 & 1.8 & 9.1 \\  
FILOU-adiabatic-high resolution & 1.59 & 0.005 & 1.6 & 8.8 \\
GYRE-nonadiabatic-low resolution & 1.68 & 0.01 & 1.2 & 8.6 \\  
\hline                  
\end{tabular}
\end{table*}

\subsection{Comparison}

We compare the model results obtained with and without the $\Delta\nu$ constraint. For the subsequent frequency difference analysis, the reference model is taken to be the minimum-misfit model obtained from the GYRE-adiabatic-high resolution configuration, fitted to 48 observed frequencies with the $\Delta\nu$ constraint. Its basic parameters are $M = 1.64,M_{\odot}$, $Z$ = 0.01, $v_{\rm rot,initial}$ = 9.6 km/s, and $\alpha_{\rm MLT}$ = 1.6. The theoretical frequencies from the other configurations are then compared with those of this reference model to evaluate the effects of spatial resolution, rotational correction, and non-adiabatic effects.

To isolate the effects of the configuration choices on the theoretical frequency differences, we fixed the stellar parameters of all models to those of the reference model. Because MESA may adopt different time steps for low and high resolution evolutionary calculations, each configuration is re-evolved with the same stellar parameters and stopped at the age of the reference model. This procedure ensures that all models are compared at the same evolutionary stage. The corresponding theoretical frequencies are then computed for each configuration.

\subsubsection{Theoretical frequency differences across different effects}

As described in Sect. 5, our work employs several pulsation codes to compute the oscillation frequencies of V1790 Ori. Because these codes differ in their numerical methods and their specific implementations, the theoretical frequencies they produce may exhibit certain deviations. Therefore, our first step is to perform a detailed comparison of the theoretical frequencies calculated by each configuration. As discussed in Section 6.1.2, the minimum-misfit model obtained from fitting 48 observed frequencies with the GYRE-adiabatic-high resolution configuration is adopted as our reference model. Therefore, in this section, we compare the theoretical frequencies with those predicted by other configuration combinations.

Table~\ref{diff1} lists the first 48 theoretical frequencies obtained from the different configurations, their differences relative to the reference case, and the corresponding mode labels $(n, \ell, m)$ assigned to the closest theoretical matches. The mode identifications are broadly consistent across the configurations, with discrepancies found for eight observed frequencies: 260.672, 464.456, 302.948, 294.924, 754.733, 801.633, 365.484, and 756.628 ($\mu$Hz). We compute the root mean square (RMS) values using two frequency sets: the 40 modes with consistent ($n,\ell,m$) identifications across all configurations, and the full set of 48 frequencies. The corresponding results are listed in the last two rows of the table.

It can be seen that, for the 40 modes with consistent $(n, \ell, m)$ identifications, the RMS$_{40}$ differences are 0.442 $\mu$Hz between low and high resolution, 0.062 $\mu$Hz between adiabatic and non-adiabatic effects, and 2.962 $\mu$Hz between the GYRE and FILOU calculations. Thus, within the consistently identified mode set, the non-adiabatic correction has the smallest direct effect on the theoretical frequencies, the change in spatial resolution produces a small but measurable difference, and the FILOU calculation gives the largest frequency shift.

When all 48 fitted frequencies are included, the corresponding RMS$_{48}$ values increase to 1.033 $\mu$Hz, 2.326 $\mu$Hz, and 3.931 $\mu$Hz, respectively. The FILOU calculation still shows the largest overall difference, while the RMS difference in the non-adiabatic case is also clearly larger than that obtained from the 40 modes with consistent identifications. This indicates that the additional eight frequencies with inconsistent mode identifications include several frequencies with relatively large deviations, thereby significantly amplifying the overall frequency differences, especially in the non-adiabatic and FILOU calculations.

\subsubsection{Observed vs theoretical frequency differences across different effects}

In the previous section, we compare the theoretical frequencies computed using different pulsation codes to assess the impact of code-specific numerical implementations and approximations. However, the ultimate goal is to evaluate how well these theoretical models reproduce the observed frequencies. Therefore, in this section, we compare the theoretical frequencies with the observed ones. The results are summarized in Table~\ref{diff2}, which lists the 48 theoretical frequencies for each configuration along with their differences relative to the observed values, and the corresponding mode labels $(n, \ell, m)$ assigned to the closest theoretical matches. Here, we compute the root mean square (RMS) values using two frequency sets: the 40 modes with consistent ($n,\ell,m$) identifications across all configurations, and the full set of 48 frequencies. The corresponding results are listed in the last two rows of the table.

For the 40 modes with consistent $(n,\ell,m)$ identifications, the RMS$_{40}$ residual decreases slightly from 4.457 $\mu$Hz in the low-resolution model to 4.387 $\mu$Hz in the high-resolution model, indicating that increasing the spatial resolution leads to a small but measurable improvement in the agreement with the observed frequencies. The RMS$_{40}$ values are almost identical, 4.387 $\mu$Hz for the adiabatic calculation and 4.381 $\mu$Hz for the non-adiabatic calculation, suggesting that non-adiabatic effects have only a very limited direct impact on the fitted frequencies. The FILOU calculation, including second-order rotational corrections, gives a larger RMS$_{40}$ residual (5.331 $\mu$Hz) than the GYRE calculation (4.387 $\mu$Hz).

When all 48 fitted frequencies are included, the RMS$_{48}$ residual decreases slightly from 4.715 $\mu$Hz in the low-resolution model to 4.682 $\mu$Hz in the high-resolution model, showing that increasing the spatial resolution produces only a minor improvement in the overall agreement with the observed frequencies. RMS$_{48}$ values are 4.682 $\mu$Hz for the adiabatic case and 4.673 $\mu$Hz for the non-adiabatic case, indicating that non-adiabatic effects have a very limited direct impact on the residuals. The FILOU calculation, which includes second-order rotational corrections, gives an RMS$_{48}$ residual of 5.270 $\mu$Hz, larger than the 4.682 $\mu$Hz obtained from the GYRE calculation.

RMS$_{40}$ is used to quantify the frequency shifts for modes with consistent mode identifications, whereas RMS$_{48}$ describes the residual level obtained from the full 48-frequency comparison, including modes whose labels change across different configurations. A comparison of these two RMS estimates shows that, for the three GYRE calculations, RMS$_{48}$ is slightly larger than RMS$_{40}$. This indicates that the additional eight frequencies with inconsistent mode identifications slightly increase the overall observed-minus-theoretical residuals. Therefore, to assess the robustness of the mode identification more clearly and to perform a quantitative comparison on a consistent basis, we use in the following discussion only the RMS values computed from the 40 modes that have the same $(n, \ell, m)$ labels across all configurations.

It should be noted that, the purpose of this comparison is not to establish a final and unique mode identification for all observed frequencies. Instead, we use the observed-theoretical frequency differences as a diagnostic of how the inferred mode labels depend on the adopted modelling strategy. This distinction is important because the closest theoretical match can change when the pulsation code, spatial resolution, rotational correction, or non-adiabatic effects are varied.

Most of the observed frequencies are associated with the same ($n, \ell, m$) labels in all configurations, indicating that their identifications are relatively insensitive to the numerical setup considered here. However, eight frequencies show inconsistent mode labels among the configurations. These cases are particularly informative because they reveal where the present modelling is most sensitive to the adopted approach. They therefore provide a useful guide for deciding which strategy should be used in future, more targeted mode identification work.

The frequency at 260.672 $\mu$Hz is the most important example. This frequency was identified as the fundamental radial mode by \cite{Bedding2020} and \cite{Murphy2023}. In our comparison, the assigned label changes with the adopted configuration: the GYRE-adiabatic-high resolution and GYRE-nonadiabatic-high resolution calculations favour a non-radial ($\ell$ = 1) identification, whereas the GYRE-adiabatic-low resolution and FILOU-adiabatic-high resolution calculations associate the same observed frequency with the fundamental radial mode. In this work, the present minimum-misfit solution should be interpreted as a preferred identification within a specific modelling framework rather than as a definitive rejection of the fundamental radial mode interpretation.

This ambiguity matters because the identification of the fundamental radial mode affects the radial-order sequence and therefore the seismic constraints on the stellar mean density and evolutionary state. As shown by the comparison among the configurations in Table~\ref{table:mmisfit-model}, the inferred $M$, $\alpha_{\mathrm{MLT}}$, and $v_{\rm rot, initial}$ vary appreciably with changes in the mesh resolution, the treatment of rotational corrections, and the inclusion of non-adiabatic effects. These variations may in turn influence the preferred mode identification. Consequently, the identification of the 260.672 $\mu$Hz peak as a non-radial mode should be regarded as the preferred solution within the present modelling framework, rather than as a definitive conclusion. This peak should therefore be treated as a key target for future controlled tests. In particular, targeted high-resolution model grids constructed within the favoured parameter ranges identified in this study will be needed to quantify these systematic effects, assess the robustness of the proposed mode identification, and improve the precision and reliability of the inferred stellar parameters.

In addition, FILOU gives eight frequencies with absolute differences smaller than 1 $\mu$Hz, compared with seven for the GYRE-adiabatic-high resolution case, six for the GYRE-adiabatic-low resolution case, and six for the GYRE-nonadiabatic-high resolution case. At the same time, however, FILOU also produces several comparatively large deviations, including a maximum absolute difference of 17.002 $\mu$Hz. This mixed behaviour shows that the table alone cannot determine which configuration provides the best fit. Instead, the local agreement with individual frequencies should be considered together with the robustness of the mode identifications.

Overall, Table~\ref{table:mmisfit-model} and Table~\ref{diff2} show that the choice of modelling strategy may affect a small but physically important subset of mode identifications, most notably the candidate fundamental radial mode. The main implication of the present comparison is therefore to motivate future modelling focused on testing the robustness of the mode labels, rather than to identify any one of the current configurations as providing a uniquely best global mode identification.

\section{Discussion}
\subsection{Large separation}

\cite{Bedding2020} calculate the large separation of $\Delta\nu$ $\sim$ 87 $\mu$Hz for V1790 Ori. They use the Python package échelle \citep{Hey2020} to interactively fine-tune $\Delta\nu$ in the échelle diagram until the highest-frequency radial mode is aligned into a vertical ridge. In this process, they use supposedly radial modes with orders ranging from $n$ = 4 to 8.

\cite{Murphy2023} re-derive $\Delta\nu$ of V1790 Ori. In their analysis, the stellar oscillation frequencies are expressed as \citep{White2011}:

\begin{equation}
\nu = \Delta \nu (n + \ell/2 + \epsilon)   \label{equation1}
\end{equation}

and this relation is used to determine $\Delta\nu$ for their model frequencies. Specifically, they perform a linear regression fit to the guessed radial mode frequencies from $n$ = 5 to 9, yielding a slope of $\Delta\nu$ and a y-intercept of $\epsilon$. Finally, they obtain the large separation for V1790 Ori to be $\Delta\nu$ $\sim$ 89 $\mu$Hz.

\begin{figure*}[t]
   \includegraphics[width=\textwidth]{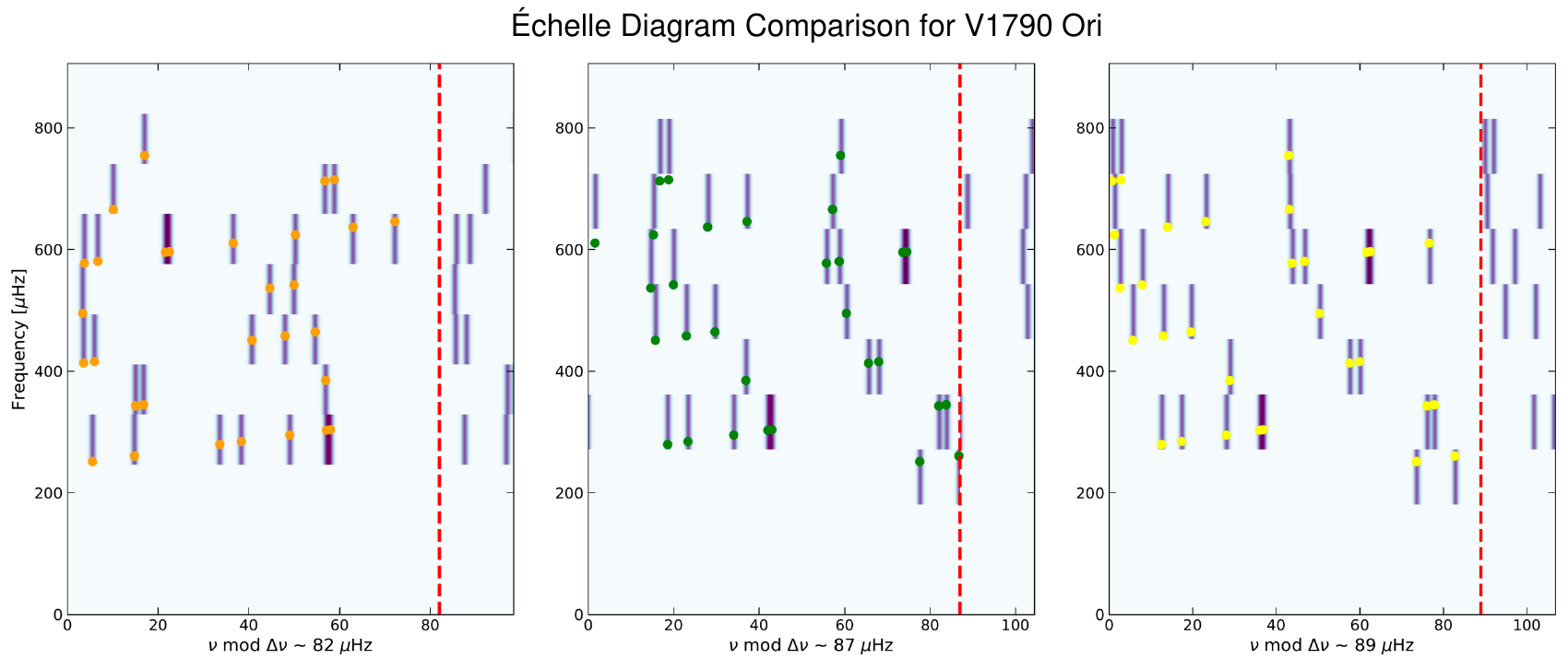}
      \caption{Échelle diagrams of V1790 Ori constructed using three different large separations. From left to right, the $\Delta\nu$ values are adopted from this work, \cite{Bedding2020} and \cite{Murphy2023}. The purple bars represent 30 observed frequencies with the highest amplitudes (normalized in the plot). The red dashed lines indicate the adopted $\Delta\nu$, while the orange, green and yellow symbols also mark observed frequencies.}
         \label{3ED}
   \end{figure*}

In this work, we obtain 82 $\mu$Hz as $\Delta\nu$ for V1790 Ori (see Sect. 3). To facilitate comparison, échelle diagrams of V1790 Ori are constructed using these three different large separations, as shown in Fig.~\ref{3ED}. Comparing these three cases allows us to directly assess how different $\Delta \nu$ choices affect the interpretation of the ridge structure. 

The comparison between this work and those of \cite{Bedding2020} and \cite{Murphy2023} is intended to be discussed only in the context of complementary approaches. The three studies emphasize different aspects of the information content, focusing respectively on visual alignment, model-based analysis, and statistical extraction. \cite{Bedding2020} determine $\Delta\nu$ by interactively adjusting the échelle diagram so that the guessed $\ell$ = 0 ridge appeared vertical. This approach inevitably involves a degree of subjectivity due to its reliance on visual inspection. \cite{Murphy2023}, on the other hand, obtain $\Delta\nu$ by fitting model frequencies with Eq.~\ref{equation1}. However, this equation is oversimplified and  based on an asymptotic relation originally derived for solar-type stars, whereas $\delta$ Scuti stars are known not to oscillate in the asymptotic regime (e.g. \citealt{AGH2009,Mirouh2017}). In this work, we analyse the observational data directly using three independent mathematical techniques, rather than relying on visual inspection or fitting models. Therefore, we contend that $\Delta \nu$ = 82 $\mu$Hz is the most robust value for V1790 Ori.

\subsection{Model with or without large separation constraints}

We analyse the parameter distributions of candidate models with and without a $\Delta \nu$ constraint. For the selection of candidate models, the range of N is determined mainly from the stable regions in the $\chi^2$/N -- N diagrams. In the case without the $\Delta \nu$ constraint, this stable region corresponds to N $\sim$ 55 -- 62 (see Sect. 6.1.1). When the $\Delta \nu$ constraint is included, the corresponding stable region shifts to N $\sim$ 43 -- 51 (see Sect. 6.1.2). Therefore, we do not impose the same N range in the two cases. Instead, the candidate models are selected separately under the two sets of constraints, requiring both a sufficiently large number of matched frequencies and a relatively low value of $\chi^2$/N.

For the selection of the fitting quality range, we adopt a relative threshold in the normalised statistic $\chi^2$/N, in order to make the selection criteria more comparable between the two cases. Specifically, we require $\chi^2$/N $\leq$ $(\chi^2/N)_{\rm min}$ (1+$\epsilon$). In the case without the $\Delta \nu$ constraint, the selected candidate models have $\chi^2$/N $\simeq$ 0.13 -- 0.14. The upper end of the selected range corresponds to a relative tolerance of approximately $\epsilon \simeq 0.08$ with respect to the minimum value of $\chi^2$/N. We therefore adopt the same relative tolerance for the case with the $\Delta \nu$ constraint. Since the minimum value of $\chi^2$/N in this case is approximately 0.45, this gives an upper limit of about 0.49. The selected candidate models with the $\Delta \nu$ constraint have $\chi^2$/N $\simeq$ 0.45 -- 0.49.

It should be emphasised that the numerical values of $\chi^2$/N for the two model sets do not necessarily represent the same fitting quality, because both the number of matched frequencies and the objective function are different when the $\Delta \nu$ constraint is included. The comparison is therefore based on the normalised statistic $\chi^2$/N together with a similar relative tolerance around the minimum value obtained under each set of constraints. Consequently, the subsequent comparison focuses on how the $\Delta \nu$ constraint affects the range and concentration of the parameter distributions, as well as the model degeneracy, rather than on a direct comparison of the $\chi^2$/N values themselves.

\begin{figure}
   \centering
   \includegraphics[width=\hsize]{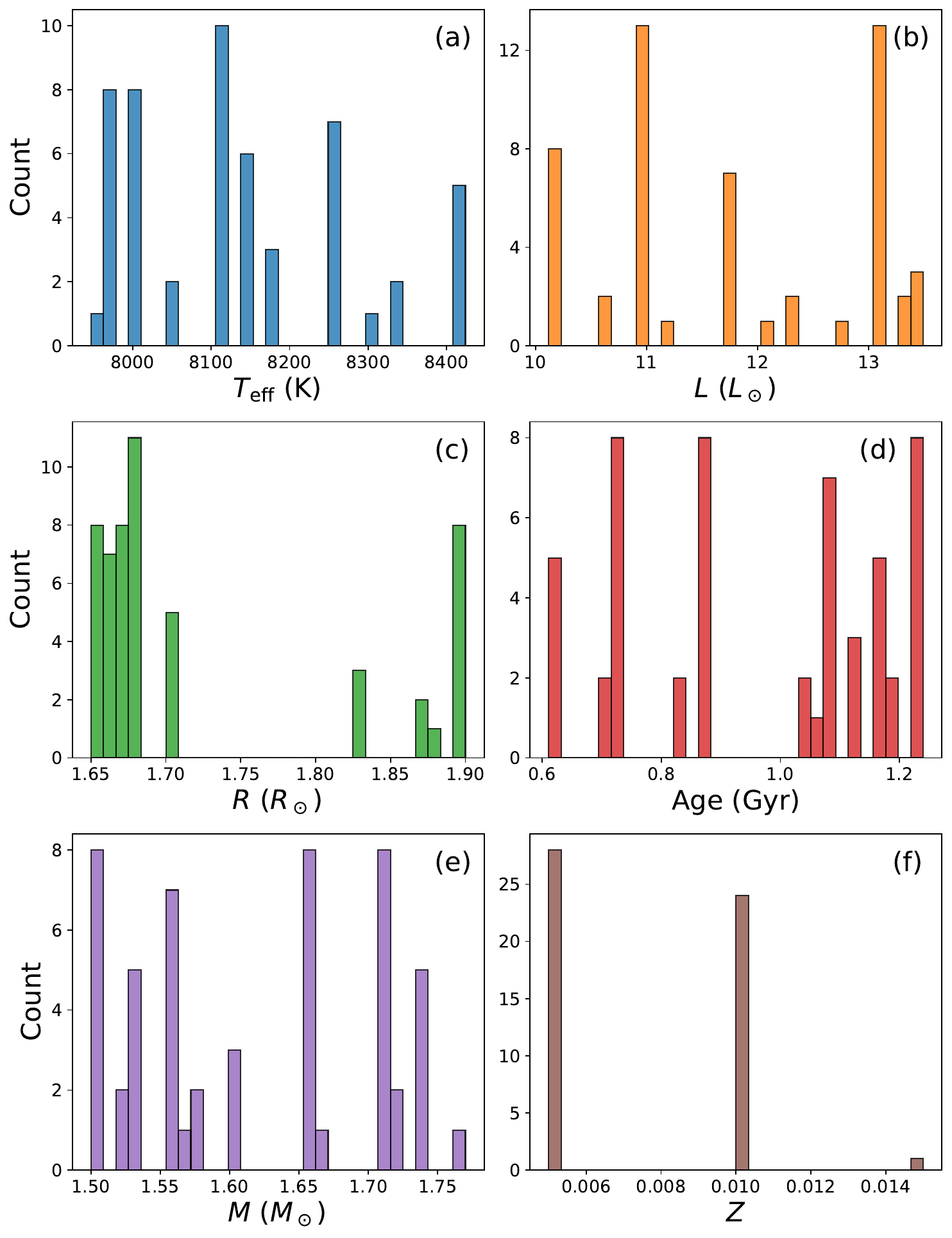}
      \caption{Histograms of the candidate model parameter distributions ($T_{\mathrm{eff}}$, $L$, $R$, age, $M$ and $Z$ in panels a, b, c, d, e, f, respectively). Models computed using GYRE-adiabatic-low resolution combination, without large separation constraint, for N = 55 – 62 and $\chi^2$/N values between 0.13 and 0.14. The height of each bar represents the number of models retained at each parameter value.}
         \label{model_parameter}
   \end{figure}

\begin{figure}
   \centering
   \includegraphics[width=\hsize]{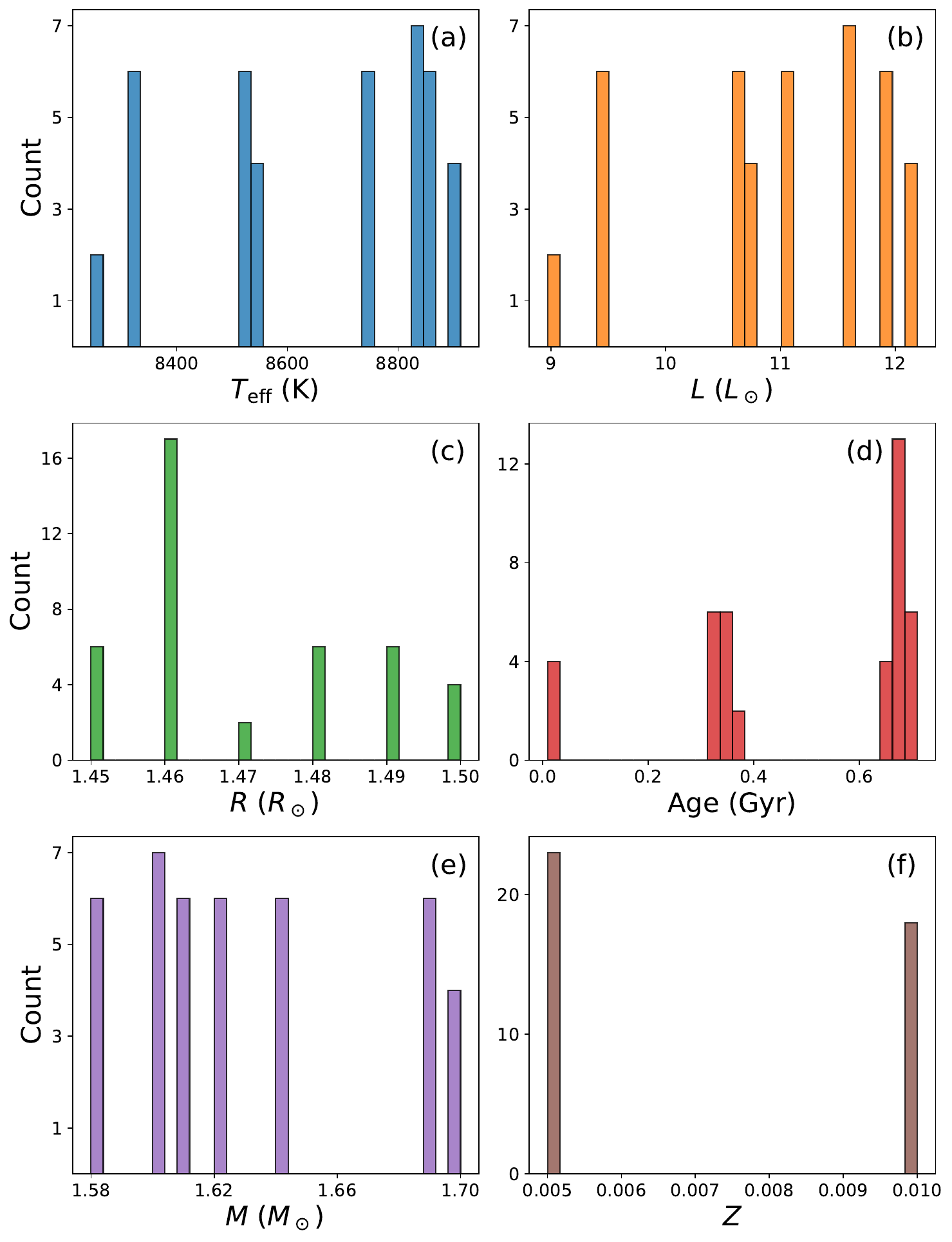}
      \caption{Histograms of the candidate model parameter distributions ($T_{\mathrm{eff}}$, $L$, $R$, age, $M$ and $Z$ in panels a, b, c, d, e, f, respectively). Models computed using GYRE-adiabatic-high resolution combination, with large separation constraint, for N = 43 – 51 and $\chi^2$/N values between 0.45 and 0.49. The height of each bar represents the number of models retained at each parameter value.}
         \label{model_parameter_dnu}
   \end{figure}

We examine the distributions of effective temperature $T_{\rm eff}$, luminosity $L$, stellar radius $R$, age, mass $M$ and metallicity $Z$ for the selected models. Fig.~\ref{model_parameter} and Fig.~\ref{model_parameter_dnu} present the corresponding histograms for the models obtained without and with the $\Delta\nu$ constraint, respectively, where the height of each bar indicates the number of models at a given parameter value.

As shown in these figures, before applying the $\Delta \nu$ constraint, the candidate models exhibit broad distributions in several fundamental parameters, indicating a significant degeneracy when only classical observational constraints are used. This is particularly evident for the radius, age, and mass: the radius extends from about 1.65 $R_{\mathrm{\odot}}$ to nearly 1.90 $R_{\mathrm{\odot}}$, the age spans approximately 0.6 -- 1.25 Gyr, and the mass ranges from about 1.50 $M_{\mathrm{\odot}}$ to 1.75 $M_{\mathrm{\odot}}$. These broad parameter ranges suggest that models with different evolutionary stages, masses, and metallicities can simultaneously satisfy the basic observational constraints, resulting in a relatively dispersed set of candidate models.

After imposing the $\Delta \nu$ constraint, the parameter space of the candidate models becomes significantly narrower and is shifted with respect to the unconstrained case. The most pronounced changes are found in the radius and age distributions. The radius decreases from a broad range of about 1.65 -- 1.90 $R_{\mathrm{\odot}}$ to a much narrower interval of approximately 1.45 -- 1.50 $R_{\mathrm{\odot}}$. This indicates that $\Delta \nu$, which is closely related to the mean stellar density, provides a strong constraint and efficiently excludes models with larger radii and lower mean densities. At the same time, the age distribution shifts towards younger models, from about 0.6 -- 1.25 Gyr to mainly 0 -- 0.7 Gyr. This suggests that the models satisfying the $\Delta \nu$ constraint preferentially correspond to earlier evolutionary stages.

The mass distribution also becomes narrower after imposing the $\Delta \nu$ constraint, decreasing from approximately 1.50 -- 1.75 $M_{\mathrm{\odot}}$ to about 1.58 -- 1.70 $M_{\mathrm{\odot}}$. Although the reduction in the mass range is less pronounced than that in the radius, the low-mass models and part of the high-mass models are excluded. This indicates that the $\Delta \nu$ constraint not only restricts the stellar radius, but also further constrains the allowed mass radius combinations through its dependence on the mean stellar density. In contrast, the luminosity distribution changes more moderately. After applying the $\Delta \nu$ constraint, the luminosity still covers approximately 9 -- 12 $L_{\mathrm{\odot}}$, although the distribution is slightly shifted towards lower luminosities compared with the unconstrained case. This behaviour is related to the significant decrease in radius: even with a higher effective temperature, a smaller radius can still lead to a comparable or slightly lower luminosity.  

The effective temperature distribution shifts overall towards higher values after imposing the $\Delta \nu$ constraint. Without the $\Delta \nu$ constraint, the models are mainly located in the range of about 8000 -- 8400 K, whereas after applying this constraint, the candidate models extend to approximately 8300 -- 8900 K. This indicates that the models satisfying the $\Delta \nu$ constraint tend to correspond to hotter, smaller radius, and younger stellar structures.

Regarding the metallicity, both sets of models are mainly concentrated around $Z$ $\sim$ 0.005 and $Z$ $\sim$ 0.010. Without the $\Delta \nu$ constraint, a small number of models with higher metallicity, for example $Z$ $\sim$ 0.015, are also present. After imposing the $\Delta \nu$ constraint, these high metallicity models are largely excluded, and the remaining candidate models are mainly concentrated around the two lower metallicity values. This indicates that the 
$\Delta \nu$ constraint also provides some filtering of the metallicity, although its effect is less direct and less pronounced than for the radius and age.

Overall, the $\Delta \nu$ constraint significantly improves the efficiency of model selection, especially by reducing the uncertainties in radius, age, and mass. After imposing this constraint, the candidate models evolve from a relatively dispersed set of evolutionary states into a more confined group of younger, hotter, and smaller radius models with a narrower mass range. This demonstrates that $\Delta \nu$, as a seismic diagnostic, can effectively reduce the degeneracy caused by classical observational constraints and plays a key role in determining the fundamental stellar parameters and evolutionary stage.

\subsection{Frequency differences}

In this section, we analyse the theoretical frequency differences between various effects, together with the deviations between theoretical and observed frequencies, to investigate how different physical effects impact the pulsation modes. As discussed in Sects. 6.2.1 and 6.2.2, we adopt the RMS frequency differences computed from the 40 modes with consistent $(n,\ell,m)$ identifications as the final frequency offsets (i.e., $\mathrm{RMS}_{40}$), thereby avoiding the conflation of genuine frequency shifts with changes in mode identification.

\subsubsection{Frequency differences between first and second order rotational corrections}

In this work, the first and second order rotational corrections of the pulsation modes are computed using GYRE and FILOU, respectively. As shown in Table~\ref{diff1}, under high resolution and adiabatic conditions, GYRE and FILOU frequencies differ by about 2.962 $\mu$Hz. These discrepancies arise because the perturbative treatment of GYRE has certain limitations. GYRE first calculates the eigenfrequencies $\sigma_{\mathrm{0}}$ for a non-rotating model, and then applies the first-order approximation to introduce rotational frequency shifts for modes with different azimuthal orders. The rotational shift is expressed as \citep{Aerts2010}:

\begin{equation}
\Delta  \sigma \approx m(1-C_{nl}) \Omega   \label{equation5}
\end{equation}

 Here, $\sigma$ is the angular frequency in the inertial frame, $m$ is the azimuthal order, $C_{nl}$ is the Ledoux constant \citep{Ledoux1951} and $\Omega$ is the angular rotation frequency of the star. Since the frequencies of $p$ modes are generally much higher than the stellar rotation frequency, the Coriolis force acts only as a small perturbation. For slowly rotating stars, the $p$-mode splitting is generally uniform and can be well described by the first-order term. However, for rapidly rotating stars, higher-order rotational effects (e.g. $\Omega^2$) introduce additional frequency perturbations \citep{Saio1981,Dziembowski1992}, such as asymmetric frequency splittings \citep{Guo2024} and mode coupling, which are not included in GYRE’s computations. Moreover, Rapid rotation causes centrifugal deformation, breaking the star’s spherical symmetry, invalidating perturbative methods, and complicating pulsation mode calculations (e.g. \citealt{Reese2021,Mirouh2022}). 

For FILOU, it includes second-order rotational corrections and more detailed treatment of splitting and coupling. In FILOU, oscillation calculations use pseudo-rotating models, accounting for rotation via a correction to the spherically symmetric component of centrifugal acceleration \citep{Suarez2008}. FILOU’s procedure has three main steps: (1) compute zeroth-order eigenfrequencies from the equilibrium model; (2) apply second-order rotational corrections to obtain rotation-modified frequencies; (3) identify near-degenerate mode pairs and calculate their corrections \citep{Suarez2002,Suarez2006a}. This hierarchical approach allows FILOU to accurately model the effects of shellular rotation on frequencies and eigenfunctions. Moreover, differences in numerical methods and boundary conditions between the codes also contribute to the deviations. In fact, this has already been done in the ESTA series of comparisons \citep{Moya2008}, and a similar analysis was also carried out by \cite{Li2025}.
 
The difference of approximately 2.962 $\mu$Hz can significantly affect mode identification and constraints on stellar structure and physical parameters in precise frequency fitting. This highlights that, in the study of pulsations in moderately rotating stars, selecting an appropriate pulsation code and cross-validating results knowing the strengths and limitations of different methods is crucial for improving the accuracy of theoretical frequencies.

Furthermore, as shown in Table~\ref{diff2}, the $\mathrm{RMS}_{40}$ differences between the theoretical frequencies and the observed frequencies are approximately 4.387 $\mu$Hz for GYRE and 5.331 $\mu$Hz for FILOU. This does not necessarily imply that the second order rotational correction gives a poorer physical description. Rather, the comparison is performed using the minimum-misfit model selected from the GYRE-adiabatic-high resolution configuration, and the stellar parameters are kept fixed when the other configurations are evaluated. FILOU and GYRE use the same set of matched modes and both account for rotational splitting, but FILOU includes second-order rotational corrections and near-degeneracy effects. These effects can modify both the frequencies and the character of coupled modes, making the resulting frequency residuals more sensitive to small changes in the stellar model and to the mode-matching procedure.

The larger $\mathrm{RMS}_{40}$ residual obtained with FILOU therefore suggests that a model optimised under the GYRE configuration may not remain optimal when second-order rotational corrections and mode coupling are included. In future modelling, FILOU calculations should be performed on a denser parameter grid, allowing the best-fitting model to be re-optimised within the same frequency computation framework. In addition to refining the main stellar parameters, it may also be useful to explore additional degrees of freedom, such as the initial helium abundance ($Y$) or selected metal abundances, since these quantities affect the internal structure and hence the oscillation spectrum. Such an approach would allow a more robust assessment of the combined effects of rotational corrections, mode coupling, and chemical composition on the theoretical frequencies.

\subsubsection{Frequency differences between low and high resolution}

In this work, we construct rotating stellar models using MESA, employing both low ($\sim$ 800 mesh points) and high ($\sim$ 4000 mesh points) resolution to assess the impact of spatial resolution on stellar structure and the resulting pulsation frequencies. We then compute the corresponding theoretical frequencies for the two sets of models and compared them directly. The results are summarized in Table~\ref{diff1}.

The $\mathrm{RMS}_{40}$ difference between frequencies computed at low and high resolution models is about 0.442 $\mu$Hz. This indicates that spatial resolution has a measurable effect on the computed frequencies, consistent with \cite{Moya2008} and \cite{Li2025}. Increasing the number of mesh points allows the stellar interior to be resolved more finely. As a result, small changes in the equilibrium structure, especially in regions with steep structural gradients or in the propagation cavities of the modes, can lead to noticeable differences in the computed frequencies. Such differences may become important for precise mode identification and for frequency fitting based on high precision space photometry.

We further compared the theoretical frequencies obtained from the low and high resolution models with the observed frequencies, as shown in Table~\ref{diff2}. The $\mathrm{RMS}_{40}$ residuals between the theoretical and observed frequencies are 4.457 $\mu$Hz for low resolution model and 4.387 $\mu$Hz for high resolution model. The high resolution model therefore gives a slightly smaller global frequency residual. Although the improvement is modest, this comparison suggests that insufficient spatial resolution can introduce additional numerical offsets in the computed mode frequencies.

Overall, these results show that the spatial mesh resolution affects the numerical accuracy of the theoretical frequencies. For the present model, this effect is smaller than the overall theoretical--observed frequency residuals, but it is not negligible compared with the level of precision required for detailed mode identification. This sensitivity therefore motivates the use of sufficiently high numerical resolution and, where necessary, a denser grid in the stellar parameter space to properly account for these effects. High resolution stellar models are thus preferable when individual theoretical and observed frequencies are compared, especially in cases where small changes in the internal structure may affect the mode frequencies.

\subsubsection{Frequency differences between adiabatic and non-adiabatic effects}

Stellar rotation and pulsation involve energy transport, and previous studies (e.g., \citealt{Moya2003,Montalban2007}) have shown that pulsation properties are highly sensitive to non-adiabatic effects, which are also crucial for mode stability and excitation \citep{Houdek2015,Sun2024}. Therefore, we compute both adiabatic and non-adiabatic frequencies with GYRE to assess the impact of non-adiabatic effects.

For the high-resolution models, we compare the adiabatic and non-adiabatic mode frequencies computed with GYRE. The $\mathrm{RMS}_{40}$ difference between the two sets of theoretical frequencies is about 0.062 $\mu$Hz, indicating that the non-adiabatic treatment introduces only a small correction to the mode frequencies for the present model and mode sample.

This conclusion is further supported by the comparison with the observed frequencies. The $\mathrm{RMS}_{40}$ residuals between the theoretical and observed frequencies are 4.387 $\mu$Hz for adiabatic calculation and 4.381 $\mu$Hz for non-adiabatic calculation. The improvement is therefore only 0.006 $\mu$Hz, which is negligible compared with the overall theoretical -- observed frequency discrepancy. Thus, although non-adiabatic effects produce a measurable change in the theoretical frequencies, they do not substantially improve the frequency matching in this case.

This suggests that the remaining frequency residuals are not mainly caused by the adiabatic approximation itself, but are more likely associated with other modelling factors, such as the stellar structure, rotational treatment, mode identification, model parameter selection, or numerical resolution. Non-adiabatic calculations nevertheless remain essential for studying mode stability, excitation, and damping. The present comparison only indicates that their direct effect on the fitted mode frequencies is limited for the present model.

\section{Summary}

Using TESS data, we detect a total of 69 significant pulsation frequencies for V1790 Ori. Based on these modes, the large frequency separation $\Delta \nu$ is determined to be approximately 82 $\mu$Hz, with three techniques: the Fourier transform (FT), the autocorrelation function (AC) and the histogram of frequency differences (HFD). We compute rotating MESA models at two spatial resolutions and calculate
theoretical frequencies with GYRE, including adiabatic and non-adiabatic effects with first-order rotational corrections, and with FILOU, including second-order rotational corrections.

When constraining the minimum-misfit models, we consider both cases with and without $\Delta \nu$. Our results indicate that for stars with a large number of pulsation modes, incorporating $\Delta \nu$ to guide the fitting between observed and theoretical frequencies can be an effective approach for correct mode identification. Moreover, introducing the large frequency separation as a fitting variable significantly improves the efficiency of model selection. It can effectively reduce the parameter degeneracy caused by relying solely on classical observational constraints and plays a key role in determining the fundamental stellar parameters and evolutionary stage. Our minimum-misfit model obtained from the GYRE-adiabatic-high resolution configuration gives the stellar parameters for V1790 Ori: $M$ = 1.64 $M_{\mathrm{\odot}}$, $Z$ = 0.01, log$T_{\mathrm{eff}}$ = 3.92, log$L$ = 0.97, log$R$ = 0.17. 

Taking the minimum-misfit model as the reference model, we quantified the frequency offsets introduced by the other configurations relative to this model. For the 40 modes with stable $(n,\ell,m)$ labels, the RMS$_{40}$ frequency offsets are 0.442 $\mu$Hz between low and high spatial resolution, 0.062 $\mu$Hz between adiabatic and non-adiabatic effects, and 2.962 $\mu$Hz between GYRE and FILOU calculations. When all 48 fitted frequencies are included, the corresponding RMS$_{48}$ values are 1.033, 2.326, and 3.931 $\mu$Hz. The comparison between RMS$_{40}$ and RMS$_{48}$ indicates that second-order rotational corrections have the largest impact on the theoretical frequencies, while also showing that a small number of modes with unstable identifications contribute significantly to the full frequency differences.

We then compared the theoretical frequencies with the observed frequencies. For the 40 stable-label modes, increasing the structural resolution slightly reduces the RMS$_{40}$ residual from 4.457 to 4.387 $\mu$Hz, while the non-adiabatic effects further changes it only marginally from 4.387 to 4.381 $\mu$Hz. The FILOU calculation gives a larger RMS$_{40}$ residual: from 4.387 to 5.331 $\mu$Hz. For the full 48-frequency comparison, the corresponding RMS$_{48}$ residuals are 4.715, 4.682, 4.673, and 5.270 $\mu$Hz. The FILOU result should not be interpreted as evidence that second-order rotational corrections are physically less appropriate; rather, it indicates that the model optimised within the GYRE framework may not remain optimal once second-order rotational corrections and near-degeneracy effects are included. A denser model grid and re-optimisation within the FILOU framework are therefore required for a fairer assessment.

Several observed frequencies retain stable mode labels across all configurations, but a small subset is sensitive to the adopted modelling strategy. The most important case is the 260.672 $\mu$Hz peak, previously identified as the fundamental radial mode. In the present comparison, its assigned mode label changes between configurations, so its identification should be regarded as model dependent rather than definitive. Future targeted high resolution modelling should therefore focus on testing the robustness of these sensitive mode identifications and on quantifying systematic uncertainties caused by rotational treatment, structural resolution, chemical composition, and numerical implementation.

Overall, the present analysis shows that $\Delta \nu$ is essential for obtaining physically meaningful model selection and reducing parameter degeneracy in V1790 Ori. Among the frequency computation effects tested here, the second order rotational is the dominant source of theoretical frequency shifts, spatial resolution produces smaller but non-negligible changes, and non-adiabatic effects have only a limited direct impact on the fitted mode frequencies for the present model. The remaining frequency residuals are therefore unlikely to be caused mainly by the adiabatic approximation itself, but are more likely associated with other modelling factors, including stellar structure, rotational treatment, mode identification, model parameter selection, or numerical resolution. Non-adiabatic calculations nevertheless remain essential for studying mode stability, excitation, and damping.

In addition, inter-code comparisons with recently developed oscillation tools, such as StORM \citep{Vanlaer2025}, may provide an important next step to assess systematic uncertainties associated with different numerical implementations and rotational treatments.

\begin{acknowledgements}
      The authors acknowledge Javier Pascual Granado for his valuable help in MultiModes.
      This research is supported by the National Natural Science Foundation of China (grant Nos. 12473043, 12003020, 12433013) and Shaanxi Fundamental Science Research Project for Mathematics and Physics (Grant No. 23JSY015).
      XYS would like to acknowledge the funding from the China Scholarship Council scholarship (No 202406280278).
      Zhaoyu Zuo acknowledges support from the China Scholarship Council (CSC).
      AGH, JCS and GMM acknowledge support from the project PID2023-149439NB-C43 funded by MICIU/AEI/10.13039/501100011033 and by FEDER, UE.
      GMM acknowledges financial support from Junta de Andalucia through the program Emergia (EMEC$\_$2023$\_$00533).
      The data presented in this paper were obtained from the Mikulski Archive for Space Telescopes (MAST).
\end{acknowledgements}

\begin{appendix} 
\captionsetup[table]{position=top}  
\section{}
\begin{table}[ht]
\caption{Multi-frequency solution of the light curves of V1790 Ori.}
\label{appendix}
\centering
\begin{tabular}{cccccc}
\hline \hline
$f$ ($\mu$Hz) & A (ppt) & SNR & $f$ ($\mu$Hz) & A (ppt) & SNR \\
\hline
260.764 & 4.544 & 210.6 & 801.910 & 0.131 & 6.4 \\
450.683 & 1.869 & 88.6  & 344.051 & 0.128 & 6.2 \\
251.539 & 0.920 & 44.0  & 365.613 & 0.123 & 5.9 \\
666.146 & 0.755 & 36.2  & 317.060 & 0.121 & 5.8 \\
577.755 & 0.643 & 31.2  & 524.884 & 0.122 & 5.8 \\
303.935 & 0.579 & 28.1  & 711.516 & 0.120 & 5.8 \\
624.259 & 0.487 & 23.3  & 375.440 & 0.119 & 5.7 \\
343.021 & 0.457 & 23.3  & 560.428 & 0.116 & 5.6 \\
536.586 & 0.353 & 17.1  & 531.262 & 0.116 & 5.6 \\
464.618 & 0.356 & 17.1  & 756.898 & 0.108 & 5.3 \\
344.780 & 0.356 & 16.8  & 268.148 & 0.110 & 5.2 \\
279.560 & 0.317 & 15.8  & 509.525 & 0.106 & 5.2 \\
457.963 & 0.324 & 15.7  & 293.819 & 0.099 & 4.9 \\
284.340 & 0.314 & 15.3  & 770.081 & 0.111 & 4.8 \\
595.648 & 0.296 & 14.6  & 768.322 & 0.100 & 4.8 \\
580.718 & 0.263 & 12.7  & 857.211 & 0.095 & 4.7 \\
712.720 & 0.222 & 11.4  & 503.322 & 0.096 & 4.6 \\
303.056 & 0.237 & 10.8  & 811.528 & 0.093 & 4.6 \\
295.023 & 0.209 & 10.0  & 297.731 & 0.093 & 4.5 \\
714.803 & 0.196 & 9.4   & 281.030 & 0.094 & 4.4 \\
495.359 & 0.191 & 9.2   & 275.521 & 0.090 & 4.4 \\
541.933 & 0.182 & 8.8   & 843.819 & 0.088 & 4.3 \\
610.567 & 0.173 & 8.3   & 975.139 & 0.088 & 4.3 \\
384.884 & 0.161 & 7.9   & 660.370 & 0.090 & 4.4 \\
596.470 & 0.161 & 7.8   & 22.350  & 0.089 & 4.4 \\
415.961 & 0.152 & 7.2   & 781.146 & 0.088 & 4.3 \\
413.565 & 0.145 & 7.2   & 478.785 & 0.087 & 4.3 \\
646.134 & 0.142 & 6.9   & 854.456 & 0.090 & 4.2 \\
755.000 & 0.142 & 6.8   & 533.530 & 0.087 & 4.2 \\
636.944 & 0.141 & 6.8   & 724.468 & 0.085 & 4.2 \\
682.569 & 0.152 & 6.7   & 266.921 & 0.083 & 4.1 \\
683.947 & 0.137 & 6.6   & 330.822 & 0.083 & 4.1 \\
587.905 & 0.134 & 6.5   & 1.701   & 0.083 & 4.1 \\
229.722 & 0.134 & 6.5   & 855.440 & 0.083 & 4.1 \\
422.708 & 0.132 & 6.4   &         &       &     \\
\hline
\end{tabular}
\tablefoot{$f$: Observed frequencies obtained by MultiModes; A: Amplitude; SNR: Signal-to-Noise Ratio.}
\end{table}     

\captionsetup[table]{position=top} 
\begin{table*}[ht]
\caption{Theoretical frequency differences across different effects.}
\label{diff1}
\centering         
\begin{tabular}{ccccccc}
\hline \hline
GYRE, ad, high res & GYRE, ad, low res & Diff & GYRE, nad, high res & Diff & FILOU, ad, high res & Diff \\
\hline      
263.926 (1,1,0) & 257.461 (1,0,0) & 6.465 & 263.926 (1,1,0) & 0.000 & 256.735 (1,0,0) & 7.191 \\
447.006 (3,2,-1) & 446.952 (3,2,-1) & 0.054 & 447.010 (3,2,-1) & -0.004 & 448.529 (3,2,-1) & -1.523 \\
252.404 (1,1,-1) & 252.327 (1,1,-1) & 0.078 & 252.404 (1,1,-1) & 0.000 & 250.877 (1,1,-1) & 1.527 \\
661.830 (6,1,-1) & 661.834 (6,1,-1) & -0.004 & 661.747 (6,1,-1) & 0.083 & 658.510 (6,1,-1) & 3.320 \\
566.741 (4,2,2) & 567.616 (4,2,2) & -0.876 & 566.744 (4,2,2) & -0.003 & 560.545 (4,2,2) & 6.196 \\
308.718 (1,2,-1) & 308.609 (1,2,-1) & 0.109 & 308.718 (1,2,-1) & 0.000 & 311.844 (1,2,-1) & -3.126 \\
625.495 (5,2,0) & 625.787 (5,2,0) & -0.292 & 625.468 (5,2,0) & 0.027 & 627.331 (5,2,0) & -1.836 \\
338.678 (1,2,1) & 339.315 (1,2,1) & -0.637 & 338.678 (1,2,1) & 0.000 & 336.361 (1,2,1) & 2.316 \\
541.653 (4,2,0) & 541.923 (4,2,0) & -0.270 & 541.656 (4,2,0) & -0.003 & 543.118 (4,2,0) & -1.466 \\
472.379 (4,0,0) & 472.661 (4,0,0) & -0.281 & 472.477 (3,2,1) & -0.098 & 470.830 (3,2,1) & 1.549 \\
353.632 (1,2,2) & 354.638 (1,2,2) & -1.006 & 353.632 (1,2,2) & 0.000 & 347.739 (1,2,2) & 5.893 \\
275.445 (1,1,1) & 275.915 (1,1,1) & -0.470 & 275.445 (1,1,1) & 0.000 & 273.879 (1,1,1) & 1.566 \\
459.707 (3,2,0) & 459.960 (3,2,0) & -0.253 & 459.711 (3,2,0) & -0.004 & 460.810 (3,2,0) & -1.103 \\
283.316 (0,2,1) & 283.878 (0,2,1) & -0.562 & 283.316 (0,2,1) & 0.000 & 282.564 (0,2,1) & 0.752 \\
600.515 (5,2,-2) & 600.204 (5,2,-2) & 0.311 & 600.491 (5,2,-2) & 0.024 & 597.505 (5,2,-2) & 3.010 \\
578.352 (5,1,-1) & 578.333 (5,1,-1) & 0.019 & 578.345 (5,1,-1) & 0.007 & 575.399 (5,1,-1) & 2.953 \\
716.371 (7,0,0) & 716.702 (7,0,0) & -0.331 & 716.244 (7,0,0) & 0.127 & 715.112 (7,0,0) & 1.259 \\
299.103 (0,2,2) & 300.033 (0,2,2) & -0.929 & 299.103 (0,2,2) & 0.000 & 298.704 (1,2,-2) & 0.399 \\
293.747 (1,2,-2) & 293.264 (1,2,-2) & 0.483 & 293.747 (1,2,-2) & 0.000 & 293.665 (0,2,2) & 0.082 \\
721.712 (6,2,1) & 722.330 (6,2,1) & -0.618 & 721.540 (6,2,1) & 0.172 & 721.438 (6,2,1) & 0.274 \\
495.678 (4,1,-1) & 495.642 (4,1,-1) & 0.036 & 495.683 (4,1,-1) & -0.005 & 492.968 (4,1,-1) & 2.711 \\
550.836 (5,0,0) & 551.129 (5,0,0) & -0.293 & 550.943 (5,0,0) & -0.106 & 549.886 (5,0,0) & 0.950 \\
613.002 (5,2,-1) & 612.993 (5,2,-1) & 0.009 & 612.977 (5,2,-1) & 0.026 & 614.029 (5,2,-1) & -1.027 \\
383.391 (2,2,0) & 383.641 (2,2,0) & -0.250 & 383.392 (2,2,0) & -0.001 & 384.212 (2,2,0) & -0.822 \\
603.290 (5,1,1) & 603.878 (5,1,1) & -0.588 & 603.282 (5,1,1) & 0.008 & 599.749 (5,1,1) & 3.541 \\
413.947 (3,1,-1) & 413.898 (3,1,-1) & 0.049 & 413.949 (3,1,-1) & -0.003 & 411.457 (3,1,-1) & 2.490 \\
411.122 (2,2,2) & 412.073 (2,2,2) & -0.951 & 411.124 (2,2,2) & -0.001 & 402.700 (2,2,2) & 8.422 \\
650.503 (5,2,2) & 651.399 (5,2,2) & -0.896 & 650.474 (5,2,2) & 0.029 & 644.323 (5,2,2) & 6.181 \\
744.586 (7,1,-1) & 744.615 (7,1,-1) & -0.030 & 756.812 (7,1,0) & -12.226 & 761.057 (7,1,0) & -16.471 \\
637.995 (5,2,1) & 638.589 (5,2,1) & -0.594 & 637.967 (5,2,1) & 0.028 & 637.438 (5,2,1) & 0.557 \\
684.258 (6,2,-2) & 683.967 (6,2,-2) & 0.290 & 684.099 (6,2,-2) & 0.159 & 680.369 (6,2,-2) & 3.889 \\
686.919 (6,1,1) & 687.531 (6,1,1) & -0.612 & 686.830 (6,1,1) & 0.089 & 682.894 (6,1,1) & 4.025 \\
590.829 (5,1,0) & 591.114 (5,1,0) & -0.285 & 590.822 (5,1,0) & 0.008 & 593.594 (5,1,0) & -2.765 \\
231.041 (-1,2,0) & 230.823 (-1,2,0) & 0.218 & 231.041 (-1,2,0) & 0.000 & 233.457 (-1,2,0) & -2.416 \\
426.029 (3,1,0) & 426.271 (3,1,0) & -0.242 & 426.031 (3,1,0) & -0.003 & 427.736 (3,1,0) & -1.707 \\
400.011 (3,0,0) & 400.279 (3,0,0) & -0.268 & 400.169 (3,0,0) & -0.158 & 399.380 (3,0,0) & 0.631 \\
799.069 (8,0,0) & 799.423 (8,0,0) & -0.354 & 803.911 (7,2,1) & -4.842 & 804.474 (7,2,1) & -5.405 \\
344.500 (2,1,0) & 344.720 (2,1,0) & -0.221 & 344.500 (2,1,0) & -0.001 & 345.715 (2,1,0) & -1.216 \\
356.378 (2,2,-2) & 356.771 (2,1,1) & -0.393 & 356.380 (2,2,-2) & -0.001 & 359.252 (2,2,-2) & -2.873 \\
323.703 (1,2,0) & 323.968 (1,2,0) & -0.265 & 323.704 (1,2,0) & 0.000 & 324.266 (1,2,0) & -0.562 \\
520.292 (4,1,1) & 520.852 (4,1,1) & -0.560 & 520.296 (4,1,1) & -0.005 & 517.189 (4,1,1) & 3.102 \\
709.227 (6,2,0) & 709.542 (6,2,0) & -0.315 & 709.059 (6,2,0) & 0.167 & 711.430 (6,2,0) & -2.203 \\
369.799 (2,2,-1) & 369.714 (2,2,-1) & 0.084 & 369.800 (2,2,-1) & -0.001 & 372.541 (2,2,-1) & -2.742 \\
554.184 (4,2,1) & 554.756 (4,2,1) & -0.572 & 554.187 (4,2,1) & -0.003 & 553.224 (4,2,1) & 0.961 \\
529.144 (4,2,-1) & 529.113 (4,2,-1) & 0.030 & 529.147 (4,2,-1) & -0.003 & 530.271 (4,2,-1) & -1.128 \\
757.173 (7,1,0) & 757.506 (7,1,0) & -0.333 & 766.481 (7,2,-2) & -9.309 & 762.419 (7,2,-2) & -5.246 \\
268.669 (0,2,0) & 268.887 (0,2,0) & -0.218 & 268.669 (0,2,0) & 0.000 & 269.488 (0,2,0) & -0.819 \\
507.995 (4,1,0) & 508.258 (4,1,0) & -0.263 & 507.999 (4,1,0) & -0.005 & 510.223 (4,1,0) & -2.228 \\
\hline
$\mathrm{RMS}_{40}$ & -- & 0.442 & -- & 0.062 & -- & 2.962 \\
$\mathrm{RMS}_{48}$ & -- & 1.033 & -- & 2.326 & -- & 3.931 \\
\hline                  
\end{tabular}
\tablefoot{The minimum-misfit model fitted to 48 observed frequencies with the $\Delta\nu$ constraint under the GYRE-adiabatic-high resolution configuration is adopted as the reference case. RMS$_{40}$ denotes the root-mean-square frequency difference computed from the 40 modes with consistent $(n,\ell,m)$ labels across all configurations, while $\mathrm{RMS}_{48}$ is computed from all 48 fitted frequencies. All frequencies are in units of $\mu$Hz. Ad and nad denote adiabatic and non-adiabatic frequencies. Low and high res mean low and high spatial resolution, respectively. The column "Diff" denotes difference between the frequencies of each configuration and the reference values.}
\end{table*}

\captionsetup[table]{position=top}  
\begin{table*}[!t]
\caption{Observed vs theoretical frequency differences across different effects.}
\label{diff2}
\centering         
\begin{tabular}{ccccccccc}
\hline \hline
Obs\_fre & GYRE,ad  & Diff & GYRE,ad  & Diff & GYRE,nad  & Diff & FILOU,ad  & Diff  \\ 
&  high res &  & low res &  & high res &  & high res \\  
\hline                     
260.672 & 263.926 (1,1,0) & -3.255 & 257.461 (1,0,0) & 3.211 & 263.926 (1,1,0) & -3.255 & 256.735 (1,0,0) & 3.936 \\
450.526 & 447.006 (3,2,-1) & 3.520 & 446.952  (3,2,-1) & 3.573 & 447.010 (3,2,-1) & 3.516 & 448.529 (3,2,-1) & 1.997 \\
251.451 & 252.404 (1,1,-1) & -0.953 & 252.327 (1,1,-1) & -0.875 & 252.404 (1,1,-1) & -0.953 & 250.877 (1,1,-1) & 0.574 \\
665.911 & 661.830 (6,1,-1) & 4.082 & 661.834 (6,1,-1) & 4.078 & 661.747 (6,1,-1) & 4.165 & 658.510 (6,1,-1) & 7.402 \\
577.547 & 566.741 (4,2,2) & 10.807 & 567.616 (4,2,2) & 9.931 & 566.744 (4,2,2) & 10.804 & 560.545 (4,2,2) & 17.002 \\
303.824 & 308.718 (1,2,-1)& -4.894 & 308.609 (1,2,-1) & -4.784 & 308.718 (1,2,-1) & -4.894 & 311.844 (1,2,-1) & -8.020 \\
624.035 & 625.495 (5,2,0) & -1.460 & 625.787 (5,2,0) & -1.752 & 625.468 (5,2,0) & -1.433 & 627.331 (5,2,0) & -3.296 \\
342.899 & 338.678 (1,2,1) & 4.221 & 339.315 (1,2,1) & 3.584 & 338.678 (1,2,1) & 4.221 & 336.361 (1,2,1) & 6.538 \\
536.402 & 541.653 (4,2,0) & -5.250 & 541.923 (4,2,0) & -5.521 & 541.656 (4,2,0) & -5.253 & 543.118 (4,2,0) & -6.716 \\
464.456 & 472.379 (4,0,0) & -7.923 & 472.661 (4,0,0) & -8.204 & 472.477 (3,2,1) & -8.021 & 470.830 (3,2,1) & -6.374 \\
344.657 & 353.632 (1,2,2) & -8.975 & 354.638 (1,2,2) & -9.981 & 353.632 (1,2,2) & -8.975 & 347.739 (1,2,2) & -3.082 \\
279.466 & 275.445 (1,1,1) & 4.021 & 275.915 (1,1,1) & 3.551 & 275.445 (1,1,1) & 4.021 & 273.879 (1,1,1) & 5.587 \\
457.803 & 459.707 (3,2,0) & -1.904 & 459.960 (3,2,0) & -2.157 & 459.711 (3,2,0) & -1.908 & 460.810 (3,2,0) & -3.007 \\
284.243 & 283.316 (0,2,1) & 0.927 & 283.878 (0,2,1) & 0.365 & 283.316 (0,2,1) & 0.927 & 282.564 (0,2,1) & 1.679 \\
595.439 & 600.515 (5,2,-2) & -5.076 & 600.204 (5,2,-2) & -4.765 & 600.491 (5,2,-2) & -5.052 & 597.505 (5,2,-2) & -2.066 \\
580.512 & 578.352 (5,1,-1) & 2.160 & 578.333 (5,1,-1) & 2.179 & 578.345 (5,1,-1) & 2.167 & 575.399 (5,1,-1) & 5.113 \\
712.468 & 716.371 (7,0,0) & -3.903 & 716.702 (7,0,0) & -4.234 & 716.244 (7,0,0) & -3.776 & 715.112 (7,0,0) & -2.644 \\
302.948 & 299.103 (0,2,2) & 3.845 & 300.033 (0,2,2) & 2.915 & 299.103 (0,2,2) & 3.844 & 298.704 (1,2,-2) & 4.243 \\
294.924 & 293.747 (1,2,-2) & 1.177 & 293.264 (1,2,-2) & 1.660 & 293.747 (1,2,-2) & 1.177 & 293.665 (0,2,2) & 1.259 \\
714.557 & 721.712 (6,2,1) & -7.156 & 722.330 (6,2,1) & -7.774 & 721.540 (6,2,1) & -6.984 & 721.438 (6,2,1) & -6.881 \\
495.187 & 495.678 (4,1,-1) & -0.491 & 495.642 (4,1,-1) & -0.455 & 495.683 (4,1,-1) & -0.496 & 492.968 (4,1,-1) & 2.219 \\
541.748 & 550.836 (5,0,0) & -9.088 & 551.129 (5,0,0) & -9.381 & 550.943 (5,0,0) & -9.195 & 549.886 (5,0,0) & -8.138 \\
610.350 & 613.002 (5,2,-1) & -2.652 & 612.993 (5,2,-1) & -2.642 & 612.977 (5,2,-1) & -2.626 & 614.029 (5,2,-1) & -3.679 \\
384.745 & 383.391 (2,2,0) & 1.354 & 383.641 (2,2,0) & 1.104 & 383.392 (2,2,0) & 1.353 & 384.212 (2,2,0) & 0.532 \\
596.265 & 603.290 (5,1,1) & -7.025 & 603.878 (5,1,1) & -7.613 & 603.282 (5,1,1) & -7.017 & 599.749 (5,1,1) & -3.483 \\
415.819 & 413.947 (3,1,-1) & 1.872 & 413.898 (3,1,-1) & 1.921 & 413.949 (3,1,-1) & 1.870 & 411.457 (3,1,-1) & 4.362 \\
413.424 & 411.122 (2,2,2) & 2.302 & 412.073 (2,2,2) & 1.351 & 411.124 (2,2,2) & 2.300 & 402.700 (2,2,2) & 10.724 \\
645.912 & 650.503 (5,2,2) & -4.591 & 651.399 (5,2,2) & -5.487 & 650.474 (5,2,2) & -4.562 & 644.323 (5,2,2) & 1.589 \\
754.733 & 744.586 (7,1,-1) & 10.147 & 744.615 (7,1,-1) & 10.118 & 756.812 (7,1,0) & -2.079 & 761.057 (7,1,0) & -6.324 \\
636.724 & 637.995 (5,2,1) & -1.271 & 638.589 (5,2,1) & -1.865 & 637.967 (5,2,1) & -1.243 & 637.438 (5,2,1) & -0.714 \\
682.324 & 684.258 (6,2,-2) & -1.934 & 683.967 (6,2,-2) & -1.644 & 684.099 (6,2,-2) & -1.776 & 680.369 (6,2,-2) & 1.954 \\
683.702 & 686.919 (6,1,1) & -3.217 & 687.531 (6,1,1) & -3.829 & 686.830 (6,1,1) & -3.128 & 682.894 (6,1,1) & 0.808 \\
587.694 & 590.829 (5,1,0) & -3.135 & 591.114 (5,1,0) & -3.420 & 590.822 (5,1,0) & -3.127 & 593.594 (5,1,0) & -5.900 \\
229.645 & 231.041 (-1,2,0) & -1.397 & 230.823 (-1,2,0) & -1.178 & 231.041 (-1,2,0) & -1.397 & 233.457 (-1,2,0) & -3.812 \\
422.556 & 426.029 (3,1,0) & -3.472 & 426.271 (3,1,0) & -3.714 & 426.031 (3,1,0) & -3.475 & 427.736 (3,1,0) & -5.179 \\
399.200 & 400.011 (3,0,0) & -0.811 & 400.279 (3,0,0) & -1.080 & 400.169 (3,0,0) & -0.969 & 399.380 (3,0,0) & -0.181 \\
801.633 & 799.069 (8,0,0) & 2.563 & 799.423 (8,0,0) & 2.210 & 803.911 (7,2,1) & -2.278 & 804.474 (7,2,1) & -2.842 \\
343.934 & 344.500 (2,1,0) & -0.566 & 344.720 (2,1,0) & -0.786 & 344.500 (2,1,0) & -0.566 & 345.715 (2,1,0) & -1.781 \\
365.484 & 356.378 (2,2,-2) & 9.106 & 356.771 (2,1,1) & 8.713 & 356.380 (2,2,-2) & 9.104 & 359.252 (2,2,-2) & 6.232 \\
316.950 & 323.703 (1,2,0) & -6.753 & 323.968 (1,2,0) & -7.018 & 323.704 (1,2,0) & -6.754 & 324.266 (1,2,0) & -7.316 \\
524.705 & 520.292 (4,1,1) & 4.413 & 520.852 (4,1,1) & 3.853 & 520.296 (4,1,1) & 4.408 & 517.189 (4,1,1) & 7.515 \\
711.260 & 709.227 (6,2,0) & 2.034 & 709.542 (6,2,0) & 1.718 & 709.059 (6,2,0) & 2.201 & 711.430 (6,2,0) & -0.169 \\
375.304 & 369.799 (2,2,-1) & 5.506 & 369.714 (2,2,-1) & 5.590 & 369.800 (2,2,-1) & 5.505 & 372.541 (2,2,-1) & 2.763 \\
560.233 & 554.184 (4,2,1) & 6.049 & 554.756 (4,2,1) & 5.476 & 554.187 (4,2,1) & 6.046 & 553.224 (4,2,1) & 7.009 \\
531.072 & 529.144 (4,2,-1) & 1.928 & 529.113 (4,2,-1) & 1.959 & 529.147 (4,2,-1) & 1.925 & 530.271 (4,2,-1) & 0.801 \\
756.628 & 757.173 (7,1,0) & -0.544 & 757.506 (7,1,0) & -0.877 & 766.481 (7,2,-2) & -9.853 & 762.419 (7,2,-2) & -5.790 \\
268.053 & 268.669 (0,2,0) & -0.615 & 268.887 (0,2,0) & -0.833 & 268.669 (0,2,0) & -0.615 & 269.488 (0,2,0) & -1.434 \\
509.347 & 507.995 (4,1,0) & 1.353 & 508.258 (4,1,0) & 1.090 & 507.999 (4,1,0) & 1.348 & 510.223 (4,1,0) & -0.875 \\
\hline
$\mathrm{RMS}_{40}$ & -- & 4.387 & -- & 4.457 & -- & 4.381 & -- & 5.331 \\
$\mathrm{RMS}_{48}$ & -- & 4.682 & -- & 4.715 & -- & 4.673 & -- & 5.270 \\
\hline                  
\end{tabular}
\tablefoot{The minimum-misfit model fitted to 48 observed frequencies with the $\Delta\nu$ constraint under the GYRE-adiabatic-high resolution configuration. RMS$_{40}$ denotes the root-mean-square frequency difference computed from the 40 modes with consistent $(n,\ell,m)$ identifications across all configurations, while RMS$_{48}$ is computed from all 48 fitted frequencies. All frequencies are in units of $\mu$Hz. The column "Obs$\_$fre" corresponds to observed frequencies. Ad and nad denote adiabatic and non-adiabatic frequencies. Low and high res mean low and high spatial resolution, respectively. The column "Diff" denotes difference between the frequencies of each configuration and the observed values.}
\end{table*}

\end{appendix}

\end{document}